\documentclass[twocolumn,aps,prb,longbibliography,superscriptaddress,floatfix]{revtex4-2}

\usepackage{mathtools,amssymb,graphicx,units,bm}
\usepackage{amsmath}
\usepackage[plainpages=false,pdfpagelabels,colorlinks=true,linkcolor=red,urlcolor=PineGreen,citecolor=PineGreen,pdftitle={Title},pdfauthor={},pdfdisplaydoctitle=true,pdfduplex=DuplexFlipLongEdge]{hyperref}
\usepackage{soul} 
\usepackage[dvipsnames,usenames]{color}
\usepackage[normalem]{ulem}
\usepackage{color}
\usepackage{bbold} 
\usepackage{float}
\emergencystretch=\maxdimen
\usepackage{silence}
\usepackage{dcolumn}
\usepackage{bm}
\usepackage{physics}%
\usepackage[dvipsnames]{xcolor}
\usepackage{bbold} 
\usepackage{float}
\emergencystretch=\maxdimen
\DeclareMathOperator{\sign}{sign}

\begin{document}

\normalem

\title{Topological characterization of modified Kane-Mele-Rashba models via local spin Chern marker}

\author{Sebasti\~ao dos Anjos \surname{Sousa-J\'unior}}
\email{sebastiao.a.s.junior@gmail.com}
\affiliation{Department of Physics, University of Houston, Houston, Texas 77004}

\author{Marcus V. de S.  \surname{Ferraz}}
\affiliation{Departamento de F\'{i}sica, Universidade Federal do Piau\'{i}, 64049-550 Teresina, Piau\'{i}, Brazil}

\author{Jos\'{e} P. \surname{de Lima}}
\affiliation{Departamento de F\'{i}sica, Universidade Federal do Piau\'{i}, 64049-550 Teresina, Piau\'{i}, Brazil}

\author{Tarik P. \surname{Cysne}}
\email{tarik.cysne@gmail.com}
\affiliation{Instituto de F\'\i sica, Universidade Federal Fluminense, 24210-346 Niter\'oi RJ, Brazil}

\begin{abstract}
In this work, we use the local spin Chern marker (LSCM) recently introduced by Ba\`{u} and Marrazzo [Phys. Rev. B 110, 054203 (2024)] to analyze the real-space (${\bf r}$-space) topology of the ground-state electronic wave functions in a honeycomb structure described by three distinct models. The models considered here are characterized by strong Rashba spin-orbit interaction, which leads to non-conservation of the spin operator, i.e., $[\mathcal{H},\hat{s}_z]\neq 0$. The three spin-orbit couplings associated with the topological aspects of the models are: 1) Standard Kane-Mele coupling, 2) Sublattice-dependent Kane-Mele coupling, and 3) In-plane ($\hat{s}_y$) polarized Kane-Mele coupling. These couplings occur in graphene grown on suitable substrates and are relevant for modeling its van der Waals heterostructures. A particular topological phase diagram characterizes each of these spin-orbit interactions, and our calculations of LSCM, fully performed on ${\bf r}$-space, successfully capture its general features. We also performed a detailed analysis of the spectral properties of the energy and valence-projected spin matrix eigenvalues, which shows that both exhibit a gap that protects the marker. To complement, we examine the effect of disorder and spatial inhomogeneities on the LSCM for the different lattice models addressed in our work. Our results expand the applicability of the spin Chern number method to a class of lattice Hamiltonians with experimental relevance and may contribute to future research on the real-space topology of realistic materials.
\end{abstract}

\date{\today\ -- Version 1.0}
\maketitle

\section{Introduction}
\label{sec:introduction}

Insulating materials in a non-trivial topological phase are characterized by gapped bulk energy spectra, which are indexed by a topological invariant. This invariant remains unchanged under smooth modifications in the electronic Hamiltonian and is robust against weak disorder and electronic interactions \cite{Hasan-Kane-RevModPhys.82.3045, Rachel_ReportProgressInPhysics, PMelo-PhysRevB.108.195151, Mondaini-PhysRevB.104.195117,Gilardoni-PhysRevB.106.L161106}. Much of our knowledge about the topological nature of matter has emerged from studying systems in the thermodynamic limit with translational symmetry, i.e., in the band theory paradigm. In these systems, the crystalline Hamiltonian can be mapped into a Bloch Hamiltonian $\mathcal{H}({\bf k})$ defined in ${\bf k}$-space within the Brillouin Zone (BZ), which has the topology of a closed surface. From this perspective, topological invariants are regarded as global properties of Bloch states and are expressed through integrals of geometric quantities in ${\bf k}$-space. For instance, Chern insulators are characterized by a nonzero integer invariant known as the Chern number $C$, which can be computed by integrating the Berry curvature over the first BZ \cite{Haldane-PhysRevLett.61.2015, Thouless-PhysRevLett.49.405}. The occurrence of a Chern insulator phase requires the breaking of time-reversal symmetry and is associated with gapless boundary states that support anomalous Hall conductance. On the other hand, quantum spin Hall insulators (QSHI) are characterized by a $\mathbb{Z}_2$-index whose definition requires the presence of time-reversal symmetry \cite{Kane-Mele-PhysRevLett.95.226801, Kane-Mele-PhysRevLett.95.146802}. These topological phases are distinguished by boundary modes that enable spin-Hall conductance. 

In Ref. \cite{Prodan-PhysRevB.80.125327}, Prodan introduced the spin Chern number as an alternative topological invariant to characterize the QSHI phase. Unlike the $\mathbb{Z}_2$-invariant, the definition of the spin Chern number does not explicitly rely on the presence of time-reversal symmetry, but it does require a gap in the eigenvalue spectrum of the valence-projected spin matrix. Although such a gap is not a universal feature of spin-orbit coupled Hamiltonians, the spin Chern number is a genuine topological index. In time-reversal symmetric systems, it is widely accepted that it essentially contains the information of the $\mathbb{Z}_2$-invariant \cite{Vanderbilt-PhysRevB.85.115415}. This topological invariant was introduced within the framework of band theory (thermodynamic limit) and assumes a quantized value even for Hamiltonians where spin is not a conserved quantity, i.e., $\left[\mathcal{H}({\bf k}), \hat{s}_z\right]\neq 0$. Prodan's method has also been applied to define Chern numbers related to other observables, such as the isospin Chern number \cite{Sheng-PhysRevB.82.165104} and the orbital Chern number \cite{Cysne-PhysRevLett.126.056601, Cysne-PhysRevB.105.195421}, demonstrating the usefulness of this concept. An interesting aspect of this topological index is its direct connection to response functions, such as orbital Hall conductivity \cite{Cysne-PhysRevLett.126.056601, Cysne-PhysRevB.105.195421} and isospin Hall conductivity \cite{Sheng-PhysRevB.82.165104}.

The requirement for the existence of a BZ when defining a topological index constrains its application to pristine and uniform solids in the thermodynamic limit. In particular, a BZ cannot be defined in a finite system or in a system with a high degree of disorder. This compromises the topological characterization of systems with significant practical interest \cite{Caio-PhysRevB.106.245408, Felipe-PhysRevB.110.075421, Anderson-PhysRevB.108.245105}. To overcome these limitations, the tool of local markers was introduced. These local markers are mathematical objects that encode information about the topology of the ground state's electronic wave functions as real-space (${\bf r}$-space) quantities. Currently, markers to characterize distinct topological phases are available \cite{Hermann-PhysRevLett.132.073803, Mong-PhysRevB.100.205101, Piskunow-PhysRevResearch.2.013229, HOMarker-Kim2023, Gilardoni-PhysRevB.106.L161106,Regis2024}. For instance, Bianco and Resta introduced in Ref. \cite{Bianco-Resta-PhysRevB.84.241106} a formula for the \textit{local Chern marker} that allows the real-space indexation of Chern insulators. Extensions of this scheme to treat periodic superlattices \cite{Bau-Marrazzo-PhysRevB.109.014206} and inclusion of thermal effects \cite{Molignini2023-TempChernMarker} have been made. Local Chern markers have also been applied to study the robustness of the topological phase against disorder \cite{Chen-PhysRevB.109.094202} and the topological Anderson transition \cite{Caio-PhysRevB.109.L201102}. 

Recently, Ba\`u and Marrazzo \cite{Bau-Marrazzo-SpinChernMarker} introduced a real-space quantity called the Local Spin Chern Marker (LSCM), which combines the Bianco-Resta and Prodan methods to capture the topological information associated with the QSHI phase. The authors validate the method by using the standard Kane-Mele-Rashba model and comparing it to the universal $\mathbb{Z}_2$-local marker. In this work, we employ LSCM to analyze the topology of generalized Kane-Mele-Rashba models in real space. The models considered here exhibit a strong Rashba effect. In addition to the standard Kane-Mele coupling, we also consider two generalizations that are relevant for modeling graphene van der Waals heterostructures: the sublattice-resolved Kane-Mele and the in-plane Kane-Mele interactions. The sublattice-dependent Kane-Mele coupling occurs in graphene grown on transition metal dichalcogenides (TMDs) in the 2H structural phase \cite{Fabian-PhysRevB.92.155403, Fabian-PhysRevLett.120.156402}. On the other hand, the in-plane polarized Kane-Mele coupling occurs in graphene proximate to a low-symmetry substrate with strong spin-orbit interaction \cite{Cysne-PhysRevB.98.045407}. This latter type of interaction has also been predicted to occur in a low-symmetry structural phase of WTe$_2$ \cite{Jose-PhysRevLett.125.256603}. We obtain the topological phase diagrams of the three models mentioned above using ${\bf r}$-space calculation of LSCM in a honeycomb structure. The models studied here present very distinct and rich phase diagrams, and we show how the LSCM successfully captures their characteristics in ${\bf r}$ space. We also discuss the features of the energy and valence-projected spin spectra, showing that both exhibit a gap that protects the topological characterization of each model according to this index \cite{Prodan-PhysRevB.80.125327}. Our results expand the applicability of the spin Chern number to describe the topology of electronic Hamiltonians and may represent progress toward describing Chern markers related to other observables \cite{Cysne-PhysRevLett.126.056601, Cysne-PhysRevB.105.195421, Li-PhysRevB.109.155407, Li-PhysRevB.108.224422} and to applications in complex materials.

 
\section{Local Spin Chern Marker \label{SEC2}}  

For Hamiltonians that satisfy $\left[\mathcal{H},\hat{s}_z  \right]=0$, the z-component of electronic spin is a good quantum number. In this situation, the Hilbert space of the Hamiltonian can be separated into two decoupled sectors, each associated with an eigenvalue $s=\uparrow$ and $\downarrow$ of the spin operator $\hat{s}_z$. One can assign Chern numbers $C_{\uparrow}$ and $C_{\downarrow}$ to each subspace, and the quantity $\frac{1}{2}\left(C_{\uparrow}-C_{\downarrow} \right)$ is an integer number, with its non-zero value in a time-reversal symmetric system indicating a QSHI phase. For systems in the thermodynamic limit and with translational symmetry, one can map $\mathcal{H}$ into a Bloch Hamiltonian defined in the BZ $\mathcal{H}({\bf k})=e^{-i{\bf k}\cdot {\bf r}}\mathcal{H}e^{i{\bf k}\cdot {\bf r}}$ and express the Chern number associated to each sector of Hilbert space as a ${\bf k}$-integral of the spin Berry curvature: $C_{\uparrow,\downarrow}=(2\pi)^{-2}\int_{\rm BZ}d^2{\bf k}\Omega_{\uparrow,\downarrow}({\bf k})$.

In models where $\left[\mathcal{H} ({\bf k}),\hat{s}_z  \right]\neq 0$, the Hilbert space cannot be decoupled into two independent sectors, and the quantity $\frac{1}{2}\left( C_{\uparrow}-C_{\downarrow}\right)$ is no longer an integer. Then, a more elaborate procedure must be used to define an integral Chern number that represents information about the topology of the QSHI phase. The spin Chern number, first introduced in Ref. \cite{Sheng-PhysRevLett.97.036808} and later mathematically formalized in thermodynamic limit by Prodan in Ref. \cite{Prodan-PhysRevB.80.125327}, allows for the definition of a quantized invariant even in cases where $\left[\mathcal{H}({\bf k}),\hat{s}_z  \right]\neq 0$. The method consists of constructing the valence-band projector $\mathcal{P}({\bf k})$ and diagonalizing the ground-state projected spin matrix $M^{s_z}_{\text{v.b.}}({\bf k})=\mathcal{P}({\bf k})\hat{s}_z\mathcal{P}({\bf k})$. For time-reversal symmetric Hamiltonians the eigenvalues $\xi^{s_z}({\bf k})$ of the matrix $M^{s_z}_{\text{v.b.}}({\bf k})$ are symmetrically distributed around zero, within the interval $\xi^{s_z}({\bf k})\in [-1,1]$. Given that $\sigma=\text{sign}(\xi^{s_z}({\bf k}))$ is the sign of $\xi^{s_z}({\bf k})$ [$\sigma=+$ for positive $\xi^{s_z}({\bf k})$ and $\sigma=-$ for negative $\xi^{s_z}({\bf k})$] one defines the projector $\mathcal{P}_{\sigma}({\bf k})=\sum_{\xi({\bf k})| \sign[\xi({\bf k})]=\sigma} \ket{\phi_{\xi,{\bf k}}}\bra{\phi_{\xi,{\bf k}}}$ where $\ket{\phi_{\xi,{\bf k}}}$ are the eigenvectors of $M^{s_z}_{\text{v.b.}}({\bf k})$ with eigenvalue $\xi^{s_z}({\bf k})$. Using $\mathcal{P}_{\sigma}({\bf k})$, one can assign an integer Chern number $C_{\sigma}$ to each projected subspace, provided that there is a spectral gap $\Delta^{s_z}$ separating positive and negative branches of the spectrum $\xi^{s_z}({\bf k})$. The spin Chern number is defined by $C_s=\frac{1}{2}\left( C_+-C_-\right)$. A non-zero spin Chern number indicates the existence of a QSHI phase and, in time-reversal symmetric systems, is equivalent to the universal $\mathbb{Z}_2$-invariant \cite{Prodan-PhysRevB.80.125327}. This invariant is topologically protected by the energy spectrum gap $\Delta^{\rm E}$ and the gap $\Delta^{s_z}$. This method has been successfully applied to study the topological properties of uniform insulating materials and other systems from their band structures \cite{Yang2011-PhysRevLett.107.066602, Baokai-2DMat-Antimonene, Sheng-PhysRevB.82.165104, Cysne-PhysRevLett.126.056601, Cysne-PhysRevB.105.195421, Lai-PhysRevB.109.L140104, Wang-PhysRevB.108.245103, NatCommn-Deng-2020,Wang-PhysRevB.101.081109}. Note that we use the term `universal' for the $\mathbb{Z}_2$-invariant to indicate that it does not require the existence of a finite $\Delta^{s_z}$, a well-known issue with the spin Chern number (see Ref. \cite{Prodan-PhysRevB.80.125327} and discussion in Sec. \ref{SEC5}).

As discussed in the introduction, it is not possible to construct a BZ in finite or highly disordered/inhomogeneous systems. Instead, real-space markers must be used to quantify the topological properties of the electronic wave functions. Recently, the Ref. \cite{Bau-Marrazzo-SpinChernMarker} introduced a combination of Prodan and Bianco-Resta methods to define the LSCM,
\begin{eqnarray}
\mathfrak{C}_s({\bf r})=\frac{\mathfrak{C}_+({\bf r})-\mathfrak{C}_-({\bf r})}{2}, \label{SpinChernMarker} 
\end{eqnarray}
where,
\begin{eqnarray}
\mathfrak{C}_{\sigma}({\bf r})=2\pi \text{Im}\bra{{\bf r}}\mathcal{Q}_{\sigma}\hat{X} \mathcal{P}_{\sigma}\hat{Y} \mathcal{Q}_{\sigma} - \mathcal{P}_{\sigma}\hat{X} \mathcal{Q}_{\sigma}\hat{Y} \mathcal{P}_{\sigma} \ket{{\bf r}}. 
\label{Csigma}
\end{eqnarray}
In Eq. (\ref{Csigma}), $\hat{X}$ and $\hat{Y}$ are the unity cells position operators. Analogous to the procedure described in the previous paragraph for ${\bf k}$-space, one defines the projector into the valence subspace (occupied states) by $\mathcal{P}=\sum^{N_{\text{occ}}}_{n=1}\ket{u_n}\bra{u_n}$, where $\ket{u_n}$ is the eigenstate of lattice model Hamiltonian $\mathcal{H}$ of the system with state index $n$. Then, we construct the spin matrix projected onto the valence subspace $M^{s_z}_{\text{v.s.}}=\mathcal{P}\hat{s}_z\mathcal{P}$ and perform its diagonalization $M^{s_z}_{\text{v.s.}}\ket{\phi_{v}}=\xi_v^{s_z}\ket{\phi_{v}}$, where ($v=1,..., N_{\text{occ}}$). For time-reversal symmetric models, the $N_{\text{occ}}$ eigenvalues $\xi_v^{s_z}$ are symmetrically distributed around zero within the interval $[-1,1]$. The branches of this spectrum can be separated by the sign of their eigenvalues [$\sigma=+$ for positive $\xi_v^{s_z}$ and $\sigma=-$ for negative $\xi_v^{s_z}$] if there is a finite gap $\Delta^{s_z}$. The normalized eigenvectors of $M^{s_z}_{\rm v.s.}$, $\ket{\phi_v}$, has dimension $N_{\text{occ}}$ and can be written as $\ket{\phi_{v}} =\left( \beta_{1,v}, \beta_{2,v},..., \beta_{N_{\rm occ},v}\right)^{\text{T}} $, where T means the transpose operation. We then define the states,
\begin{eqnarray}
\ket{\psi_{v}} =\sum^{N_{\text{occ}}}_{\alpha=1} \beta_{v,\alpha} \ket{u_{\alpha}}. \label{Rotsigma}
\end{eqnarray}
in terms of the valence eigenstates $\ket{u_{\alpha}}$ of the lattice model Hamiltonian $\mathcal{H}$ of the system. With these states, the projectors used in Eq. (\ref{Csigma}) are given by
\begin{eqnarray}
\mathcal{P}_{\sigma}=\sum_{\xi^{s_z}_v|\sign [\xi^{s_z}_v]=\sigma} \ket{\psi_{v}}\bra{\psi_{v}}. \label{Psigma}
\end{eqnarray}
Finally, the complementary matrices in Eq. (\ref{Csigma}) are defined as $\mathcal{Q}_{\sigma} = \mathbb{1}- \mathcal{P}_{\sigma}$. Here, we adopt an equivalent, but slightly different, notation than Ba\`{u} and Marrazzo.
Alternatively, one could proceed with matrix multiplications, choosing a consistent basis to construct the projector, as done in Ref. \cite{Bau-Marrazzo-SpinChernMarker}. However, we chose to follow the equivalent construction described above, inspired by Ref. \cite{Yang2011-PhysRevLett.107.066602}. In Ref. \cite{Bau-Marrazzo-SpinChernMarker} authors also demonstrate that the LSCM defined in Eqs. (\ref{SpinChernMarker}-\ref{Psigma}) is compatible with the $\mathbb{Z}_2$ marker in the standard non-interacting Rashba-Kane-Mele model with sublattice potential. We briefly mention that other independent works \cite {Chen_2023, Wieder-Bradlyn-NatCommun} have proposed expressions for the LSCM similar to the one in Ref. \cite{Bau-Marrazzo-SpinChernMarker}.

In the following sections, we use the LSCM method to calculate the real-space topology of three spin-orbit coupled lattice Hamiltonians. We consider two types of boundary conditions: periodic boundary conditions (PBC) in Secs. \ref{SEC3}, \ref{SEC4} and \ref{SEC5} [following Ref. \cite{Chen-PhysRevB.107.045111}] and open boundary conditions (OBC) in Sec. \ref{SEC6} [following Ref. \cite{Bianco-Resta-PhysRevB.84.241106}]. Notably, the concept of local markers can be applied equally well to ${\bf r}$-space calculations in systems with both types of boundary conditions \cite{Bianco-Resta-PhysRevB.84.241106, Chen-PhysRevB.107.045111}, though PBC provides faster numerical convergence for markers at the central region. We mention in passing that the term supercell is also used to refer to a larger system composed of $L\times L$ primitive cells with PBC. However, we avoid this nomenclature in the present work and follow the terminology of Ref. \cite{Chen-PhysRevB.107.045111}.

\begin{figure}[t]
    \centering
    \includegraphics[width=0.9\linewidth,clip]{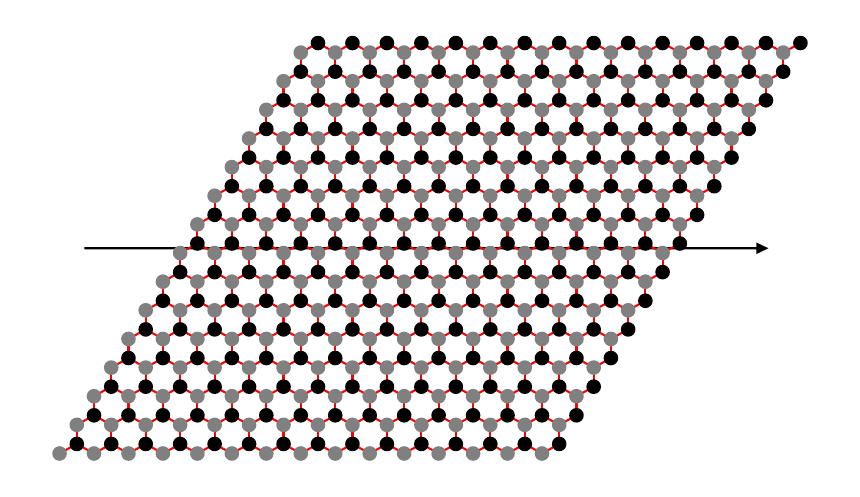}
        \caption{The typical structure used in this work to calculate LSCM. Here we show a system with linear size $L=15$. The black arrow indicates the path used in the following figures to plot the topological markers in real space. In Secs. \ref{SEC3}-\ref{SEC5}, we apply PBC to the system following Ref. \cite{Chen-PhysRevB.107.045111}. In Sec. \ref{SEC6}, we use OBC to study lateral heterostructures and the effect of disorder following Ref. \cite{Bianco-Resta-PhysRevB.84.241106}. See the discussion at the end of Sec. \ref{SEC2}.}
    \label{fig:flake}
\end{figure}


\section{Standard Kane-Mele-Rashba Hamiltonian \label{SEC3}}

\begin{figure}[t]
    \centering
    \includegraphics[width=1.0\linewidth,clip]{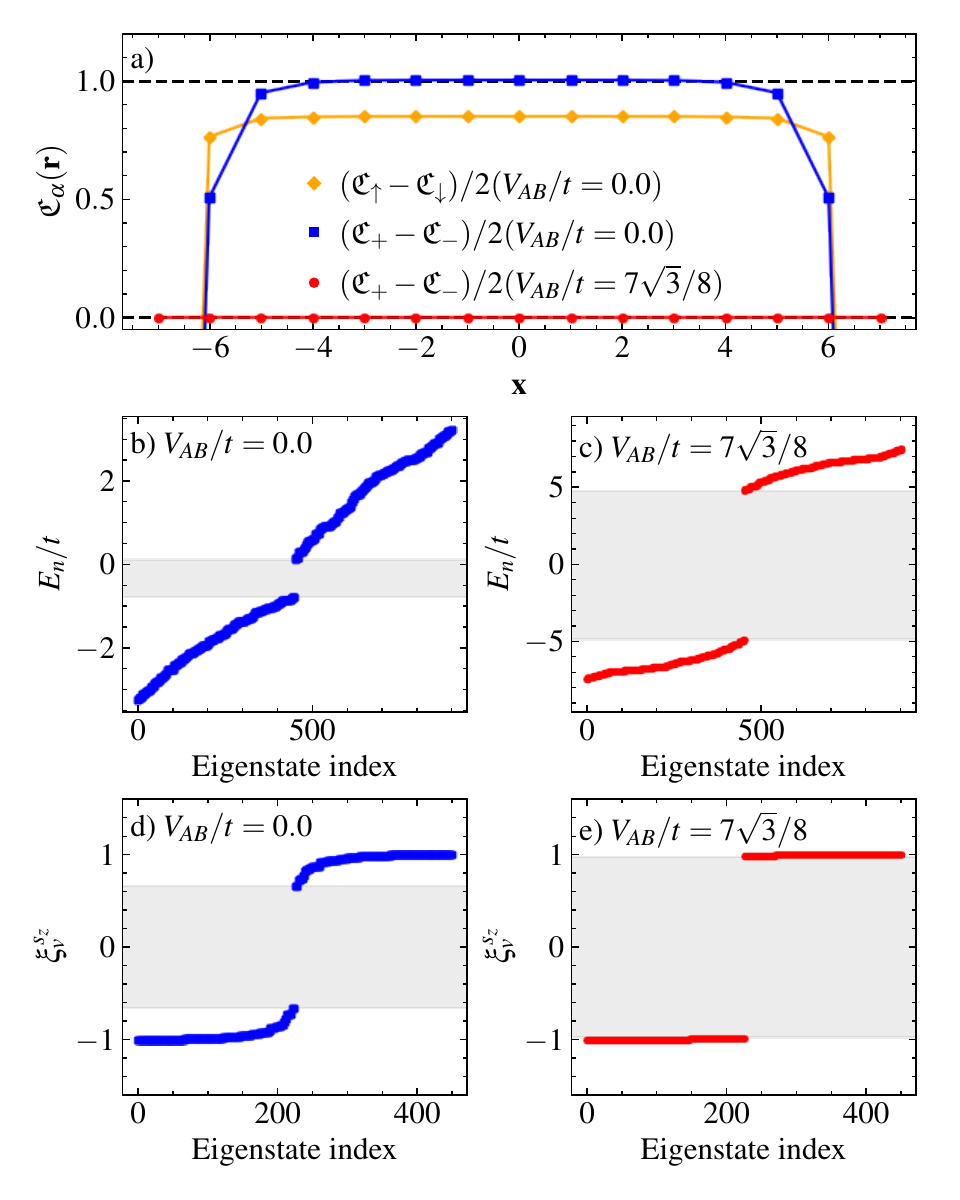}
    \caption{(a) LSCM along the ${\bf x}$-axis (black arrow in Fig. \ref{fig:flake}) for a lattice model described by the standard Kane-Mele-Rashba Hamiltonian [Eq. (\ref{SKMR})] with length $L=15$. The blue squares show the markers for a QSHI phase [$\mathfrak{C}_s({\bf 0})=1$] and the red circles for a trivial phase [$\mathfrak{C}_s({\bf 0})=0$]. We also show using yellow diamonds the values of $\left(\mathfrak{C}_{\uparrow}-\mathfrak{C}_{\downarrow}\right)/2$ for parameters in the QSHI phase that, in the presence of Rashba coupling, lead to a non-quantized quantity [see the discussion in sec. \ref{SEC2}]. (b, c) Energy spectra of the lattice model in QSHI and trivial phases, respectively. (d, e) The valence-states projected spin spectra $\xi^{s_z}_v$ of the lattice model in QSHI and trivial phases, respectively. 
    Here, we set Rashba coupling $\lambda_{\rm R}= \sqrt{3}\lambda_{\rm KM}\approx0.43 t$ and Kane-Mele coupling $\lambda_{\rm KM}=0.25 t$ in both cases, and employed PBC to lattice Hamiltonian. The sublattice potential was set $V_{\rm AB}=0$ for QSHI phase and $V_{\rm AB}=7\sqrt{3}t/8\approx 1.51 t$ in the trivial phase. The shaded rectangles indicate the gaps $\Delta^{\rm E}$ and $\Delta^{s_z}$.}
    \label{fig:Standart-Topological}
\end{figure}

\begin{figure}[t]
    \centering
    \includegraphics[width=0.95\linewidth,clip]{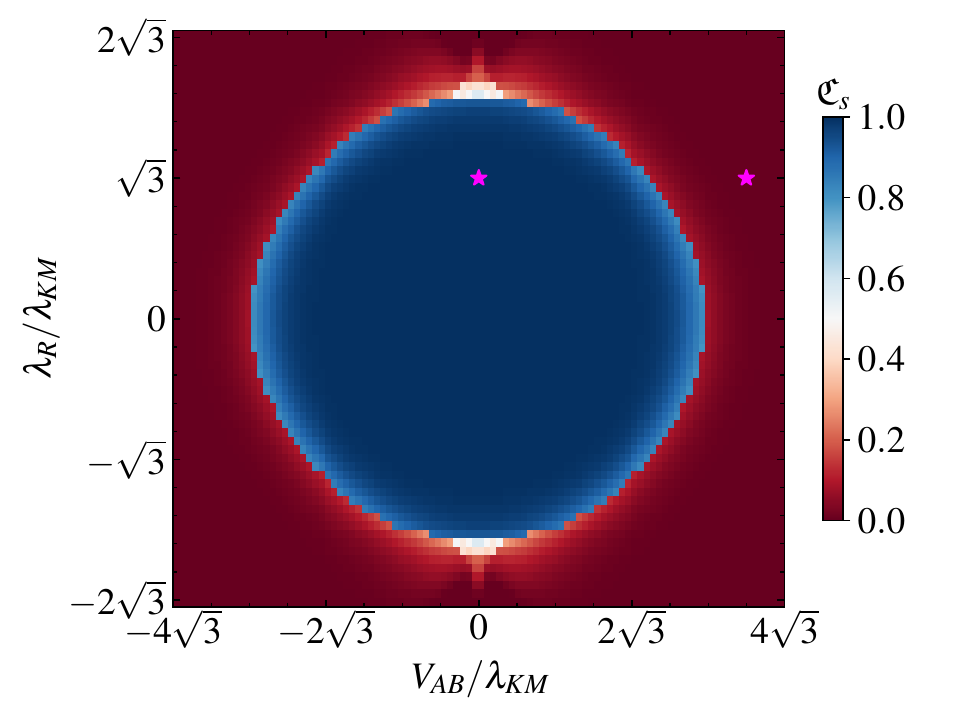}
    \caption{The LSCM phase diagram [$\mathfrak{C}_s({\bf 0})$ in Eq. (\ref{SpinChernMarker})] of the standard Kane-Mele-Rashba model [Eq. \ref{SKMR}]. Here we set $\lambda_{\rm KM}= 0.25 t$. The magenta stars represent the phase diagram points corresponding to Fig. \ref{fig:Standart-Topological}. The ${\bf r}$-space calculations were performed for a lattice Hamiltonian with $L=15$ and subjected to PBC.}
    \label{fig:pahse_kmr}
\end{figure}

In this section, we review the application of the LSCM to the standard Kane-Mele-Rashba lattice model \cite{Bau-Marrazzo-SpinChernMarker}, complementing this with an analysis of the energy and valence-projected spin spectra. The standard Kane-Mele-Rashba Hamiltonian can be cast as,
\begin{eqnarray}
\mathcal{H_{\rm std}}=\mathcal{H}_t+\mathcal{H}_{\rm AB}+\mathcal{H}_{\rm R}+\mathcal{H}_{\rm KM}. \label{SKMR}
\end{eqnarray}
We follow Ref. \cite{Chen-PhysRevB.107.045111}, considering this Hamiltonian within a lattice of linear size $L$ subjected to PBC. $\mathcal{H}_t$ represents the nearest neighbor hopping term of electrons in the honeycomb structure of Fig. \ref{fig:flake}, as given by
\begin{eqnarray}
\mathcal{H}_t=t\sum_{s=\uparrow,\downarrow}\sum_{\langle i,j \rangle}c^{\dagger}_{i,s}c_{j,s}, \label{hopping}
\end{eqnarray}
Here, $c^{\dagger}_{i,s}$ and $c_{j,s}$ are the fermionic creation and annihilation operators for electrons at sites $i$ and $j$ of the lattice and spin $s=\uparrow,\downarrow$. $\langle i,j \rangle$ indicate that the sum runs over the nearest neighbor sites. In all results presented in this work, we set the hopping amplitude $t$ as the unit of energy [$t=1$]. $\mathcal{H}_{\rm AB}$ is the sublattice potential
\begin{eqnarray}
\mathcal{H}_{\rm AB}=V_{\rm AB}\sum_{s=\uparrow,\downarrow}\sum_i\tau_ic^{\dagger}_{i,s}c_{i,s}, \label{VAB}
\end{eqnarray}
where $\tau_i=+1$ for sites $i$ belonging to sublattice A, and $\tau_i=-1$ for sites $i$ belonging to sublattice B. $\mathcal{H}_{\rm KM}$ is the standard Kane-Mele spin-orbit coupling \cite{Kane-Mele-PhysRevLett.95.146802} given by 
\begin{eqnarray}
\mathcal{H}_{\rm KM}=i\lambda_{\rm KM}\sum_{s,s'}\sum_{\langle \langle i,j\rangle \rangle} \nu_{ij} c^{\dagger}_{i,s} \left(\hat{s}_z\right)_{s,s'}c_{j,s'}.\label{szKMcoupling}
\end{eqnarray}
Here $\nu_{ij}=\text{sign}\left({\bf d}_1\times {\bf d}_2\right)_z=\pm 1$, where ${\bf d}_1$ and ${\bf d}_2$ are unit vectors along the two bonds that the electron traverses when hopping from the site $j$ to the next nearest neighbor $i$ [indicated Eq.(\ref{szKMcoupling}) by $\langle \langle i,j\rangle \rangle$]. Finally the Rashba spin-orbit coupling $\mathcal{H}_{\rm R}$ can be cast as,
\begin{eqnarray}
\mathcal{H}_{\rm R}=i\lambda_{\rm R}\sum_{s,s'}\sum_{\langle i,j\rangle}c^{\dagger}_{i,s}\left[\left( \mathbf{\hat{s}}\times\mathbf{d}_{ij} \right)_z \right]_{s,s'}c_{j,s'} \ , \label{Rashba}
\end{eqnarray}
where $\mathbf{d}_{ij}$ is the unity vector along the direction connecting site $i$ to site $j$. $\mathbf{\hat{s}}=\hat{s}_x{\bf x}+\hat{s}_y{\bf y}+\hat{s}_z{\bf z}$ is related to physical spin of electrons (up to a factor $\hbar/2$) and $\hat{s}_{x,y,z}$ are Pauli matrices. The Rashba coupling breaks $(z\rightarrow -z)$-symmetry \cite{Bychkov-Rashba_1984} and is responsible for the non-conservation of spin in the full Hamiltonian $\left[\mathcal{H}_{\rm R}, \hat{s}_z\right]\neq 0$.

We show in Fig. \ref{fig:Standart-Topological} (a) the LSCM calculated along the line indicated by the black arrow in Fig. \ref{fig:flake}. The blue squares are the LSCM for $V_{\rm AB}$ and $\lambda_{\rm R}$ adjusted to match a QSHI phase, while the red circles are the results for parameters in the trivial phase. For the system in the QSHI phase, as shown by the blue curve in Fig. \ref{fig:Standart-Topological} (a), the LSCM assumes a quantized value $+1$, except for anomalies near the edges, which also occur in usual Chern markers \cite{Bianco-Resta-PhysRevB.84.241106} even for PBC \cite{Chen-PhysRevB.107.045111}. We emphasize that this quantization of the LSCM [Eq. (\ref{SpinChernMarker})] occurs in the presence of a finite Rashba coupling, which is set to $\lambda_{\rm R}=\sqrt{3}\lambda_{\rm KM}$ in Fig. \ref{fig:Standart-Topological}. If one simply calculates $\frac{1}{2}\left(\mathfrak{C}_{\uparrow}-\mathfrak{C}_{\downarrow}\right)$, a non-quantized value is obtained, as indicated by the yellow curve in Fig. \ref{fig:Standart-Topological} (a). We present in Fig. \ref{fig:Standart-Topological} (b) and (d) the energy ($E_n$) and valence-state projected spin spectra ($\xi^{s_z}_v$) for the lattice Hamiltonian in the QSHI phase. It is worth noting that both spectra exhibit a gap, which is responsible for the topological protection of the LSCM. The red circles in Fig. \ref{fig:Standart-Topological} (a) represent the results for the marker in the topologically trivial phase, which is zero, as expected. Panels (c) and (e) of Fig. \ref{fig:Standart-Topological} show the corresponding energy and valence-projected spin spectra, which also display a gap, thereby enabling the definition of the marker. We follow Refs. \cite{Chen-PhysRevB.107.045111, Caio-PhysRevB.109.L201102} and use the central site of the lattice in Fig. \ref{fig:flake} [${\bf r}={\bf 0}$ in Eq. (\ref{SpinChernMarker})] to calculate the topological phase diagram of LSCM. We show the results for $\mathfrak{C}_s({\bf 0})$ in the $\lambda_{\rm R}$-$V_{\rm AB}$ parameter space in Fig. \ref{fig:pahse_kmr}. Setting aside the finite size effects, which are more pronounced near the topological phase transition, the phase diagram shown in Fig. \ref{fig:pahse_kmr} aligns with the one obtained using the ${\bf k}$-space formula for the $\mathbb{Z}_2$-invariant \cite{Kane-Mele-PhysRevLett.95.146802, Kane-Mele-PhysRevLett.95.226801}.


\section{Sub-lattice-dependent Kane-Mele coupling \label{SEC4}}

\begin{figure}[t]
    \centering
    \includegraphics[width=0.9\linewidth,clip]{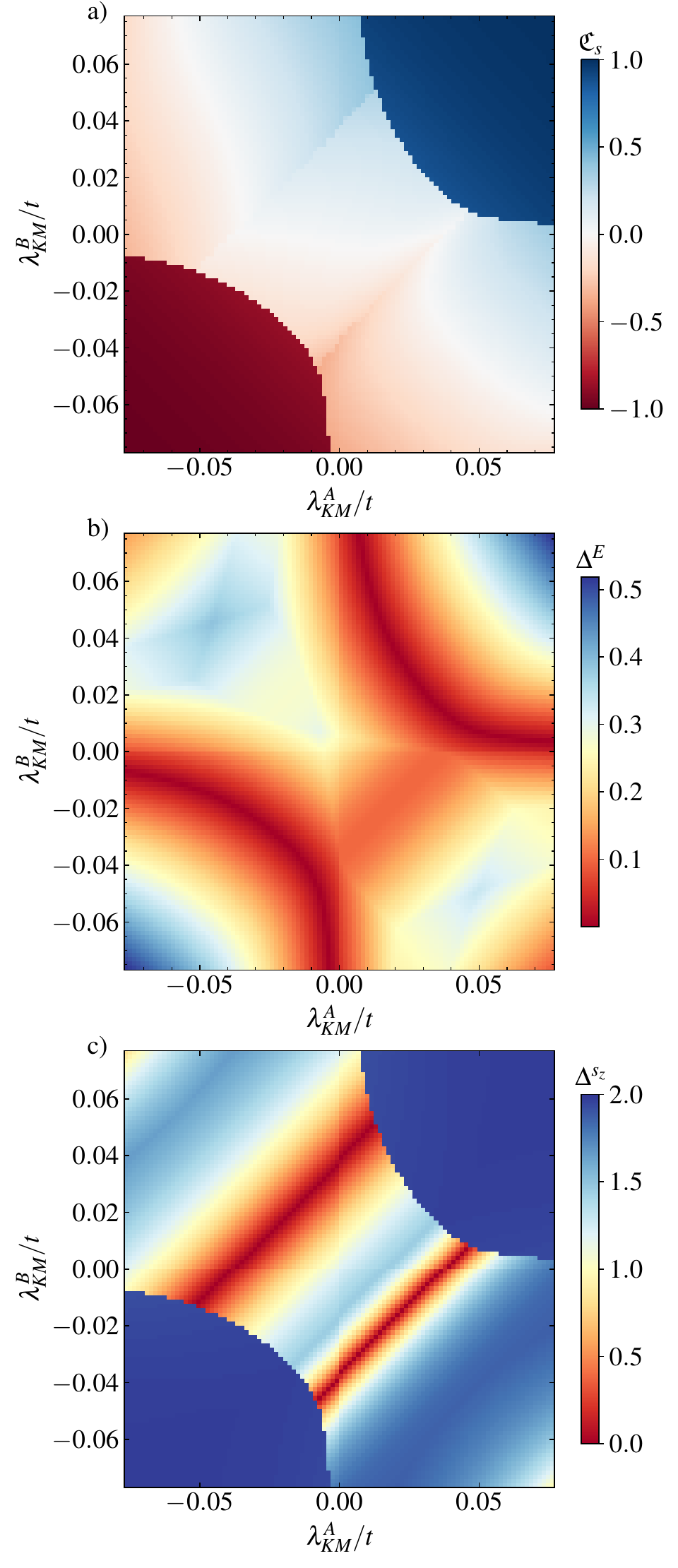}
    \caption{(a) LSCM phase diagram [$\mathfrak{C}_s({\bf 0})$ in Eq. (\ref{SpinChernMarker})] of Hamiltonian in Eq. (\ref{GEN1}). (b) Energy gap of the lattice Hamiltonian $\Delta^{\rm E}$. (c) gap of valence-projected spin spectrum $\Delta^{s_z}$. Here we set $V_{\rm AB}= 0.1t$, $\lambda_{\rm R}= 0.05t$ in a lattice with $L=15$ and subjected to PBC.}
    \label{fig:PD-sublKM}
\end{figure}

We now examine the first generalization of the standard Kane-Mele-Rashba model that we consider in this paper. When a material with strong spin-orbit coupling breaks the six-fold rotation symmetry of the honeycomb lattice reducing its point group to $C_{3v}$, a sublattice-dependent Kane-Mele spin-orbit coupling emerges \cite{Koschan-PhysRevB.95.165415, RevModPhys.92.021003},
\begin{eqnarray}
\mathcal{H}_{\rm KM(ab)}=i\sum_{s,s'}\sum_{\langle \langle i,j\rangle \rangle}\lambda^{ij}_{\rm KM} \nu_{ij} c^{\dagger}_{i,s} \left(\hat{s}_z\right)_{s,s'}c_{j,s'}.\label{szKMcouplingSubLat}
\end{eqnarray}
where $\lambda^{ij}_{\rm KM}=\lambda^{\rm A}_{\rm KM}$ for $\langle \langle i,j\rangle \rangle \in$ sublatice A, and $\lambda^{ij}_{\rm KM}=\lambda^{\rm B}_{\rm KM}$ for $\langle \langle i,j\rangle \rangle \in$ sublatice B, with $\lambda^{\rm A}_{\rm KM}\neq\lambda^{\rm B}_{\rm KM}$. This type of coupling occurs in graphene grown on TMDs in 2H structural phase, leading to intriguing spin-transport and topological features \cite{Roche-PhysRevLett.119.206601, Offidani-PhysRevLett.119.196801, Cysne-PhysRevB.97.085413, Fabian-PhysRevLett.120.156402, Power-PhysRevB.106.125417}. The complete lattice Hamiltonian considered in this section is given by 
\begin{eqnarray}
\mathcal{H_{\rm Gen 1}}=\mathcal{H}_t+\mathcal{H}_{\rm AB}+\mathcal{H}_{\rm R}+\mathcal{H}_{\rm KM(ab)}. \label{GEN1}
\end{eqnarray}
where, $\mathcal{H}_t$, $\mathcal{H}_{\rm AB}$ and $\mathcal{H}_{\rm R}$ are the terms defined in Eqs. (\ref{hopping}, \ref{VAB}) and (\ref{Rashba}). Again, we consider this Hamiltonian in the structure shown in Fig. \ref{fig:flake}, subjected to PBC. The topological phase diagram of the model in Eq. (\ref{GEN1}) was calculated in Ref. \cite{Fabian-PhysRevLett.120.156402} using the ${\bf k}$-space formula for the $\mathbb{Z}_2$ invariant. Here, we perform analogous calculations to those detailed in the previous section to index the topology of this model in real space using the LSCM defined in Eqs. (\ref{SpinChernMarker}, \ref{Csigma}). We also study their energetic and spin spectral properties.

Figure\,\ref{fig:PD-sublKM} (a) shows the topological phase diagram of the Hamiltonian in Eq. (\ref{GEN1}), based on calculations of the LSCM at the central site ($\mathfrak{C}_s({\bf 0})$) of the lattice of Fig. \ref{fig:flake}. In this figure, we set fixed values for $V_{\rm AB}=0.1 t$ and $\lambda_{\rm R}=0.05 t$ and calculated the markers for different values in $\lambda^{\rm A}_{\rm KM}-\lambda^{\rm B}_{\rm KM}$ space. The LSCM-diagram in Fig. \ref{fig:PD-sublKM} (a) matches the $\mathbb{Z}_2$-diagram in Ref. \cite{Fabian-PhysRevLett.120.156402}. In the two quadrants where $\text{sign}\left(\lambda^{\rm A}_{\rm KM}\right)=\text{sign}\left(\lambda^{\rm B}_{\rm KM}\right)$, a QSHI phase with a quantized LSCM is obtained. Notably, the LSCM in each topological region has opposite signs. The sign of the spin Chern number does not provide additional information about the topology of the ground state electronic wave function; it is simply a matter of definition in the projection procedure \cite{Prodan-PhysRevB.80.125327} described in Sec. \ref{SEC2}. The two regions correspond to the same topological order, and the connection of LSCM with the local $\mathbb{Z}_2$ marker is given by $\nu ({\bf 0})=\mathfrak{C}_s({\bf 0}) \ \text{mod} \ 2$ \cite{Bau-Marrazzo-SpinChernMarker}. The rest of the diagram is characterized by two insulating phases with zero LSCM, in agreement with Ref. \cite{Fabian-PhysRevLett.120.156402}. These phases exhibit interesting properties due to the presence of what is known as valley Zeeman coupling, including pseudo-helical states \cite{Fabian-PhysRevLett.120.156402, Costa-PhysRevLett.130.116204,dai2024twodimensionalweylnodallinesemimetal} and unconventional spin-transport phenomena \cite{Offidani-PhysRevLett.119.196801, Roche-PhysRevLett.119.206601}. The distinct regions of the phase diagram can be reached by fabricating heterostructures of graphene combined with different members of the 2H-TMD family \cite{Gmitra-PhysRevB.93.155104}. In panels (b) and (c) of Fig. \ref{fig:PD-sublKM}, we present the energy ($\Delta^{\rm E}$) and valence-projected spin-matrix ($\Delta^{s_z}$) gaps throughout the parameter space. Except in regions close to the phase transition lines, the two gaps are finite, which ensures the topological protection of LSCM in each phase.

 
\section{In-plane Kane-Mele coupling \label{SEC5}}

Now we consider a second generalization of the Kane-Mele coupling that may occur when a honeycomb lattice interacts with a low-symmetry environment. This generalization consists of a sub-lattice-dependent in-plane polarized ($\hat{s}_y$) Kane-Mele coupling 
\begin{eqnarray}
\mathcal{H}_{\rm y(ab)}=i\sum_{s,s'}\sum_{\langle \langle i,j\rangle \rangle}\lambda^{ij}_{\rm y} \nu_{ij} c^{\dagger}_{i,s} \left(\hat{s}_y\right)_{s,s'}c_{j,s'},\label{syKMcouplingSubLat}
\end{eqnarray}
and $\lambda^{ij}_{\rm y}=\lambda^{\rm A}_{\rm y}$ for $\langle \langle i,j\rangle \rangle \in$ sublatice A, and $\lambda^{ij}_{\rm y}=\lambda^{\rm B}_{\rm y}$ for $\langle \langle i,j\rangle \rangle \in$ sublatice B, with $\lambda^{\rm A}_{\rm y}\neq\lambda^{\rm B}_{\rm y}$. This effect has been predicted to occur when graphene is grown on top of materials with a low-symmetry crystal field and strong spin-orbit interaction \cite{Cysne-PhysRevB.98.045407}. A similar type of in-plane polarized coupling has also been predicted in the monolayer of WTe$_2$ in the low-symmetric 1T$_{d}$ structural phase \cite{Jose-PhysRevLett.125.256603}. The Rashba interaction tends to destroy the QSHI state in the case of $\hat{s}_{z}$-polarized Kane-Mele couplings by closing its energy gap, making it a frequently cited obstacle to observing this phase in experiments \cite{Bindel2016-NatPhys, Strom-PhysRevLett.104.256804}. On the other hand, the energy gap created by $\hat{s}_y$-Kane-Mele coupling is robust against the presence of Rashba interaction \cite{Cysne-PhysRevB.98.045407}, making the study of its properties appealing for the practical realization of topological phases. In this section, we focus on the study of the robustness of this topological phase against the Rashba effect, examining the Hamiltonian of the form:
\begin{eqnarray}
\mathcal{H_{\rm Gen 2}}=\mathcal{H}_t+\mathcal{H}_{\rm R}+\mathcal{H}_{\rm y(ab)}, \label{GEN2}
\end{eqnarray}
where, $\mathcal{H}_t$ and $\mathcal{H}_{\rm R}$ are given by Eqs. (\ref{hopping}, \ref{Rashba}).

\begin{figure}[t]
    \centering
    \includegraphics[width=1.\linewidth,clip]{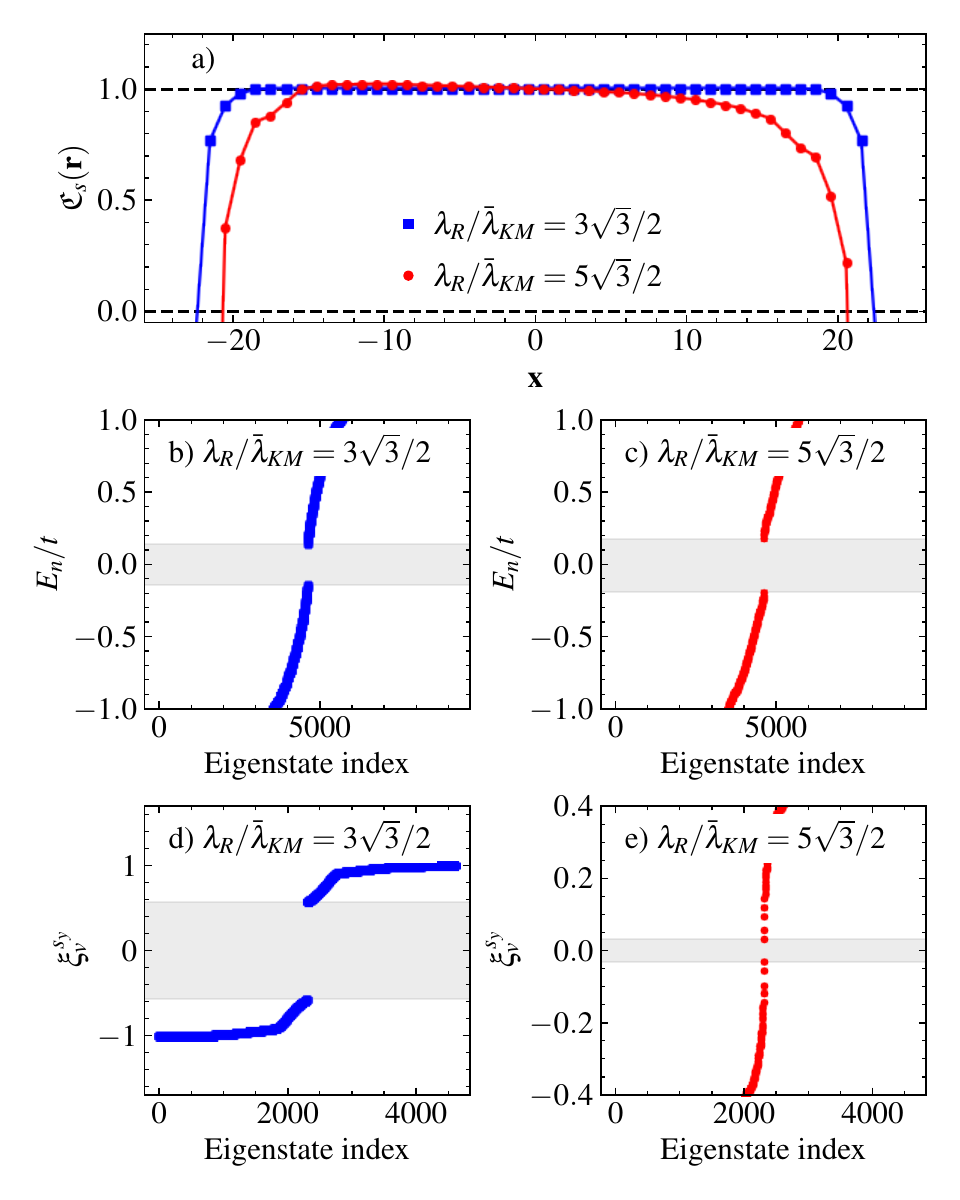}
    \caption{(a) LSCM along the ${\bf x}$-axis (black arrow in Fig. \ref{fig:flake}) for a lattice described by the model in Hamiltonian of Eq. (\ref{GEN2}) with $L=48$ and subjected to PBC. Here we set $\lambda^{\rm A}_{\rm y}= 0.19t$ and $\lambda^{\rm B}_{\rm y}= 0.21t$. The blue curve represents the results of Rashba coupling $\lambda_{\rm R}=1.5 \sqrt{3}\bar{\lambda}_{\rm KM}\approx 0.52 t$, and the red curve corresponds to $\lambda_{\rm R}=2.5 \sqrt{3}\bar{\lambda}_{\rm KM}\approx 0.87 t$, with $\bar{\lambda}_{\rm KM}=(\lambda^{\rm A}_{\rm y}+\lambda^{\rm B}_{\rm y})/2=0.2 t$. (b, c) Corresponding energy spectra. (d, e) Corresponding valence-states projected spin spectra $\xi^{s_y}_v$. The shaded rectangles indicate the gaps $\Delta^{\rm E}$ and $\Delta^{s_y}$ in both spectra.}
    \label{fig:in-plane-LSCM-Gaps}
\end{figure}

\begin{figure}[t]
    \centering
    \includegraphics[width=1.0\linewidth,clip]{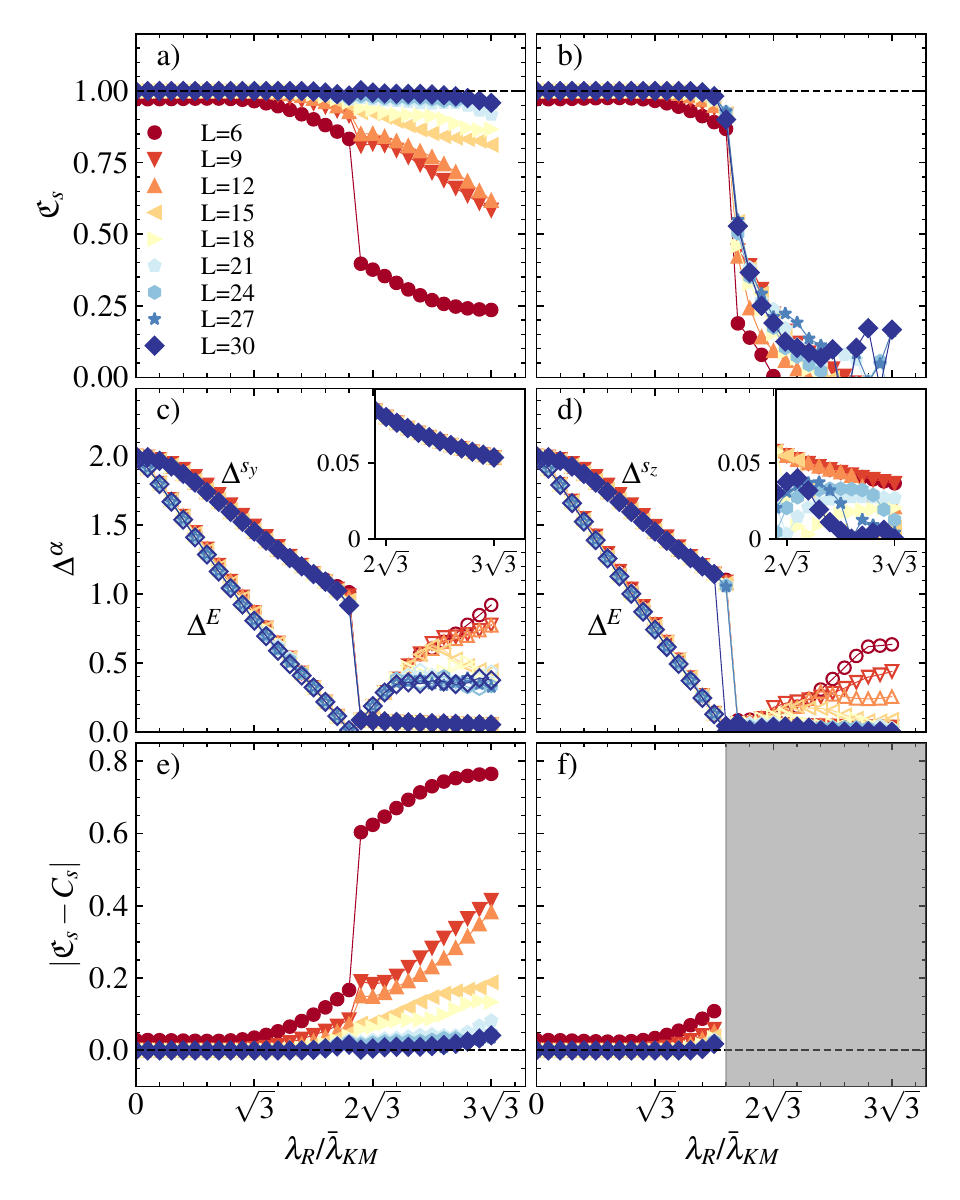}
    \caption{Upper Panels: Dependence of the LSCM $\mathfrak{C}_s({\bf 0})$ on Rashba coupling $\lambda_{\rm R}$ for lattices described by two generalized Hamiltonians: (a) $\mathcal{H}_{\rm Gen2}$ [Eq.(\ref{GEN2})] with $\lambda^{A}_{y}= 0.19 t$ and $\lambda^{B}_{y}=0.21 t$, and (b) $\mathcal{H}_{\rm Gen1}$ [Eq.(\ref{GEN1})] with $V_{\rm AB}=0$, $\lambda^{A}_{\rm KM}= 0.19 t$ and $\lambda^{B}_{\rm KM}= 0.21 t$. Central Panels: The corresponding energy and valence-projected spin-spectra gaps for the two Hamiltonian models: (c) $\Delta^{\rm E} \ [t]$ and $\Delta^{s_y}$ for $\mathcal{H}_{\rm Gen2}$, and (d) $\Delta^{\rm E} \ [t]$ and $\Delta^{s_z}$ for $\mathcal{H}_{\rm Gen1}$. The inset in each panel shows an enlarged view of region $\lambda_{\rm R}>2\sqrt{3} \bar{\lambda}_{\rm KM}$, displaying the valence-projected spin-gap results.
    Lower Panels: Analysis of the difference between LSCM $\mathfrak{C}_s({\bf 0})$ and the spin Chern number $C_s$ obtained from band theory paradigm (${\bf k}$-space formulation) for each model across different lattice sizes. The shaded region in panel (f) indicates that the energy spectra of $\mathcal{H}_{\rm Gen1}({\bf k})$ is gapless, preventing the definition of $C_s$.  Here we used lattices with size $L=6-30$ and subjected to PBC. We normalized the horizontal axis scale by setting $\bar{\lambda}_{\rm KM}=0.2 t$ in all the plots. The coupling values used in (c) and (e) for $\mathcal{H}_{\rm Gen2}$ are the same as those in (a), and the values used in (d) and (f) for $\mathcal{H}_{\rm Gen1}$ are the same as those in (b).
    }
    \label{fig:inplane-PD}
\end{figure}

The QSHI phase produced by the in-plane Kane-Mele coupling has a spin-Hall current polarized in the $y$-direction. Therefore, the projection procedure described in Sec. \ref{SEC2} must be performed using the $\hat{s}_y$ operator instead of $\hat{s}_z$. This means that, for the calculations of the LSCM for the model in Eq. (\ref{GEN2}), we obtain the states of Eq. (\ref{Rotsigma}) by diagonalizing the matrix $M^{s_y}_{\rm v.s.}=\mathcal{P}\hat{s}_y\mathcal{P}$: $M^{s_y}_{\text{v.s.}}\ket{\phi_{v}}=\xi_v^{s_y}\ket{\phi_{v}}$, where ($v=1,..., N_{\text{occ}}$). Using these states, we construct the projectors $\mathcal{P}_{\sigma}$ in Eq. (\ref{Psigma}) and apply them to the calculations of the markers in Eqs. (\ref{SpinChernMarker}) and (\ref{Csigma}). The remaining discussion follows closely in analogy to the previous case. The LSCM is protected by the gap in the energy spectrum and in the valence-projected $\hat{s}_y$-spectrum.

In Fig. \ref{fig:in-plane-LSCM-Gaps} (a), we present the LSCM calculated for the lattice model of Eq. (\ref{GEN2}) and PBC, along the line indicated by the black arrow in Fig. \ref{fig:flake}. In the results shown in Fig. \ref{fig:in-plane-LSCM-Gaps}, we use $\lambda^{\rm A}_y = 0.19 t$, $\lambda^{\rm B}_y = 0.21 t$, and two different values for $\lambda_{\rm R}$. For $\lambda_{\rm R} = 3\sqrt{3}\bar{\lambda}_{\rm KM}/2$ [$\bar{\lambda}_{\rm KM}=\left(\lambda^{\rm A}_y+\lambda^{\rm B}_y\right)/2=0.2 t$], the LSCM shows the expected profile indicative of a QSHI, which means it is quantized in the central region, with anomalies near the edges of the system due to finite-size effects. The quantization of LSCM in the central region for this case is protected by large gaps in both the energy and $\xi^{s_y}$ spectra [shaded rectangles in Fig. \ref{fig:in-plane-LSCM-Gaps} (b) and (d)]. For $\lambda_{\rm R} = 5\sqrt{3}\bar{\lambda}_{\rm KM}/2$, the finite-size effects are magnified due to the small gap $\Delta^{s_y}$ in the $\xi^{s_y}$ spectrum reported in Fig. \ref{fig:in-plane-LSCM-Gaps} (e). Nevertheless, the gap remains finite, and the central region with a quantized LSCM increases as the size of the system grows. It is worth mentioning that we used $\lambda^{\rm A}_y$ and $\lambda^{\rm B}_y$ slightly differently because, for this model, the gap in the $\xi^{s_y}$ spectrum closes when $\lambda^{\rm A}_y = \lambda^{\rm B}_y$. This is a limitation of the method, as there is no symmetry imposition that guarantees the existence of a finite gap in $\xi^{s_y}$ spectra. In this situation, the spin Chern number method fails, making it necessary to use the $\mathbb{Z}_2$ index-based marker \cite{Bau-Marrazzo-SpinChernMarker}, which depends only on the presence of a gap in the energy spectrum, to characterize the QSHI phase.

As mentioned above, an interesting feature of the spin-orbit coupling in Eq. (\ref{syKMcouplingSubLat}) is that it leads to the formation of a QSHI phase, which persists even at high Rashba coupling intensities. Figure \ref{fig:inplane-PD} (a, c) shows the LSCM at the central site $\mathfrak{C}_s({\bf 0})$ and the corresponding gaps $\Delta^{\rm E}$ and $\Delta^{s_y}$ for the lattice model in Eq. (\ref{GEN2}), as a function of $\lambda_{\rm R}/\bar{\lambda}_{\rm KM}$ [$\bar{\lambda}_{\rm KM}=0.2 t$] and different lattice sizes $L$. By inspecting the curves for $L=30$ in Fig. \ref{fig:inplane-PD} (c), one can observe two distinct behaviors in the gaps $\Delta^{\rm E}$ and $\Delta^{s_y}$. As $\lambda_{\rm R}$ increases from $0$ to a value close to $1.8 \sqrt{3}\bar{\lambda}_{\rm KM}$, the gap $\Delta^{\rm E}$ decreases from a finite value to zero, and the gap $\Delta^{s_y}$ decreases from $2$ to $\approx 1$. As the $\lambda_{\rm R}$ continues to increase beyond $1.8 \sqrt{3}\bar{\lambda}_{\rm KM}$, the gap $\Delta^{\rm E}$ reappears and then oscillates around a finite value, while the gap $\Delta^{s_y}$ jumps to a small finite value and remains nearly constant even under strong $\lambda_{\rm R}$. The finite gaps $\Delta^{\rm E}$ and $\Delta^{s_y}$ in the two regimes are reflected in a quantized LSCM at the central site across a wide range of Rashba coupling intensities. Due to the non-zero values of $\Delta^{\rm E}$ and $\Delta^{s_y}$, the LSCM is a well-defined topological invariant for all values of Rashba coupling, except at the specific value [$\lambda_{\rm R}\approx 1.8 \sqrt{3}\bar{\lambda}_{\rm KM}$] where $\Delta^{\rm E}$ vanishes. The small values of the gaps in the region where $\lambda_{\rm R}\gtrsim 1.8 \sqrt{3}\bar{\lambda}_{\rm KM}$ make the LSCM susceptible to finite-size effects. As illustrated by Fig. \ref{fig:inplane-PD} (a), quantization of LSCM is achieved only for larger lattice systems. However, the construction of projectors of Eq. (\ref{Psigma}) remains formally well-defined \cite{Prodan-PhysRevB.80.125327} despite the small values of $\Delta^{\rm E}$ and $\Delta^{s_y}$.

To contrast with the physics discussed above, we show analogous results for the model in Eq. (\ref{GEN1}) with $V_{\rm AB}=0$ in Fig. \ref{fig:inplane-PD} (b, d). Here, the projector $\mathcal{P}_{\sigma}$ is again constructed using the $\hat{s}_z$-operator, as in Sec. \ref{SEC4}. The difference in behavior for strong Rashba coupling [$\lambda_{\rm R}\gtrsim 1.6\sqrt{3}\bar{\lambda}_{\rm KM}$], compared to the case discussed in the previous paragraph, is evident. The LSCM of the central site deviates from its quantized value and strongly oscillates for all values of $L$, presenting an erratic behavior. Both gaps $\Delta^{\rm E}$ and $\Delta^{s_z}$ vanish as the size of the system increases. 

The distinct behaviors of the energy gaps shown in Fig. \ref{fig:inplane-PD} are already present in the band structure of the models in the thermodynamic limit \cite{Cysne-PhysRevB.98.045407} (${\bf k}$-space) and are reflected here in the spectra for the finite lattice system and the LSCM. In Fig. 6 (e, f), we present an analysis comparing the central-site LSCM calculated in real space for both models with the spin Chern number $C_s$ computed from the band-theory (${\bf k}$-space) paradigm for different values of $L$. It can be seen that as $L$ increases, the LSCM rapidly converges with the ${\bf k}$-space formulation.

\section{Disordered and inhomogeneous systems \label{SEC6}}

We now present some results that highlight the key advantage of using local markers to study the topology of systems. Up to this point, we have focused on homogeneous and pristine systems analyzed in real space and subjected to PBC, as in Ref. \cite{Chen-PhysRevB.107.045111}. The striking benefit of local markers lies in their ability to analyze inhomogeneous structures and systems where disorder plays a significant role. To explore the potential of LSCM fully, we switch to lattice models with OBC (also known as flakes), making the translational symmetry breaking in the systems evident \cite{Bianco-Resta-PhysRevB.84.241106}.

\subsection{Disordered systems \label{secVIA}}

\begin{figure}[t]
    \centering
    \includegraphics[width=1.\linewidth,clip]{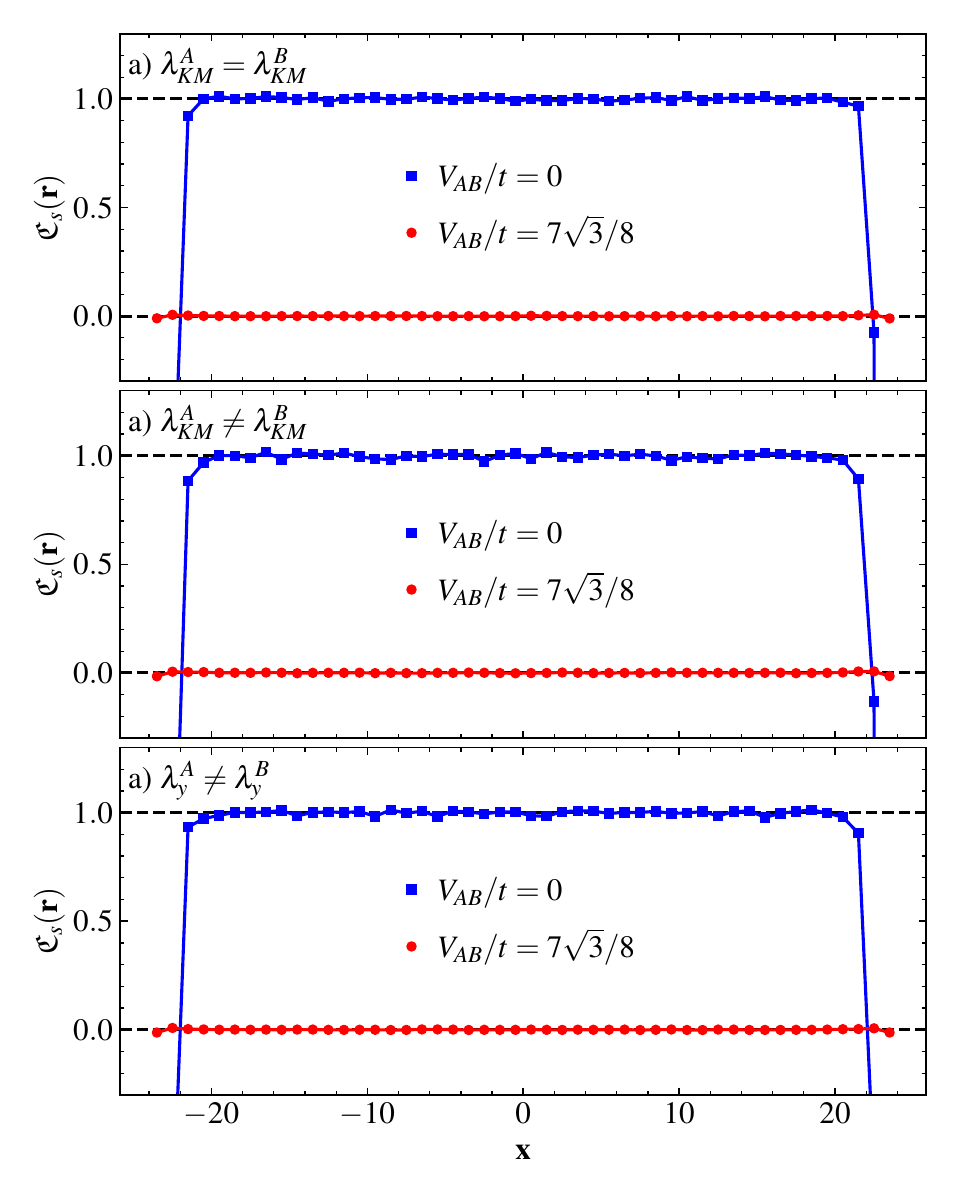}
    \caption{LSCM along the ${\bf x}$-direction for a disordered system with $L = 48$ under OBC (a flake). Each panel shows results for one of the three distinct lattice Hamiltonians studied in this work, using two sets of coupling parameters. In all panels, we introduced a configuration of disorder with strength $W = (\lambda^{A}_{\alpha}+\lambda^{B}_{\alpha})\sqrt{3}/2$. The couplings used in each panel are: (a) for the model in Eq. (\ref{SKMR}), $\mathcal{H}_{\rm std}$, we used $(V_{\rm AB}/t, \ \lambda_{\rm R}/t, \ \lambda_{\rm KM}/t)=(0.0 \ , \ 0.43, \ 0.25)$ [blue curve] and $=(1.51, \ 0.43, \ 0.25)$ [red curve]. (b) for the model in Eq. (\ref{GEN1}), $\mathcal{H}_{\rm Gen1}$, we used $(V_{\rm AB}/t, \ \lambda_{\rm R}/t, \ \lambda^{\rm A}_{\rm KM}/t, \ \lambda^{\rm B}_{\rm KM}/t)=(0.0,\ 0.35,\ 0.19,\ 0.21)$ [blue curve] and $=( 1.51,\ 0.35,\ 0.19,\ 0.21)$ [red curve]. (c) for the model in Eq. (\ref{GEN2}) added to sublattice potential [Eq. (\ref{VAB})], $\mathcal{H}_{\rm Gen2}+\mathcal{H}_{\rm AB}$, we used $(V_{\rm AB}/t,\ \lambda_{\rm R}/t,\ \lambda^{\rm A}_{\rm y}/t,\ \lambda^{\rm B}_{\rm y}/t)=( 0,\ 0.35,\ 0.19,\ 0.21)$ [blue curve] and $=(1.51,\ 0.35,\ 0.19,\ 0.21)$ [red curve]. Our implementation follows Ref. \cite{Bianco-Resta-PhysRevB.84.241106} by introducing an on-site disorder energy given by: $\epsilon = W\sum_{s=\uparrow,\downarrow} \sum_i\gamma(i)\tau_ic^{\dagger}_{i,s}c_{i,s} $, where $\gamma(i)$ is a random number uniformly distributed between $0$ and $1$. In panels (a) and (b), the LSCM was calculated using $\mathcal{P}_{\sigma}$ in Eq. (\ref{Psigma}) constructed from $\hat{s}_z$-operator [see Secs. \ref{SEC3}, \ref{SEC4}], while in panel (c), it was constructed using the $\hat{s}_y$-operator [see Sec. \ref{SEC5}].}
    \label{fig:disorder}
\end{figure}

In Fig. \ref{fig:disorder}, we present typical LSCM results for disordered samples in systems described by the three lattice Hamiltonians studied in this work. For each Hamiltonian, we select two sets of coupling constants: one placing the system in a trivial phase and the other in a topological phase. Our implementation of the disorder potential follows the one used in Ref. \cite{Bianco-Resta-PhysRevB.84.241106}. More sophisticated implementations and analyses are possible \cite{Caio-PhysRevB.109.L201102, Chen-PhysRevB.109.094202}, but we defer them to future work. In the presence of weak disorder, the LSCM fluctuates around the quantized value in the central region of the sample. More drastic disorder implementations and stronger potentials require configurational and spatial averaging to extract the quantitative agreement of local markers with the expected quantized values \cite{Caio-PhysRevB.109.L201102}. But here, we intend simply to qualitatively explore the effect of a perturbing random potential, as done in Ref. \cite{Bianco-Resta-PhysRevB.84.241106}. Finally, we pointed out that real-space calculations of local markers are typically susceptible to discontinuities at the extrema sites of systems, regardless of whether they are subjected to PBC or OBC. This is discussed in the Refs. \cite{Chen-PhysRevB.109.094202, Chen-PhysRevB.107.045111} and can be observed by contrasting Fig. \ref{fig:disorder} for OBC systems with those presented in previous sections that used PBC. Proposals to solve this well-known issue with local markers involve a redefinition of position operators \cite{Chen-PhysRevB.109.094202, Aligia-PhysRevB.107.075153, Aligia-PhysRevLett.82.2560}.

\subsection{Lateral heterojunction \label{secVIB}}

\begin{figure}
    \centering
    \includegraphics[width=1.\linewidth,clip]{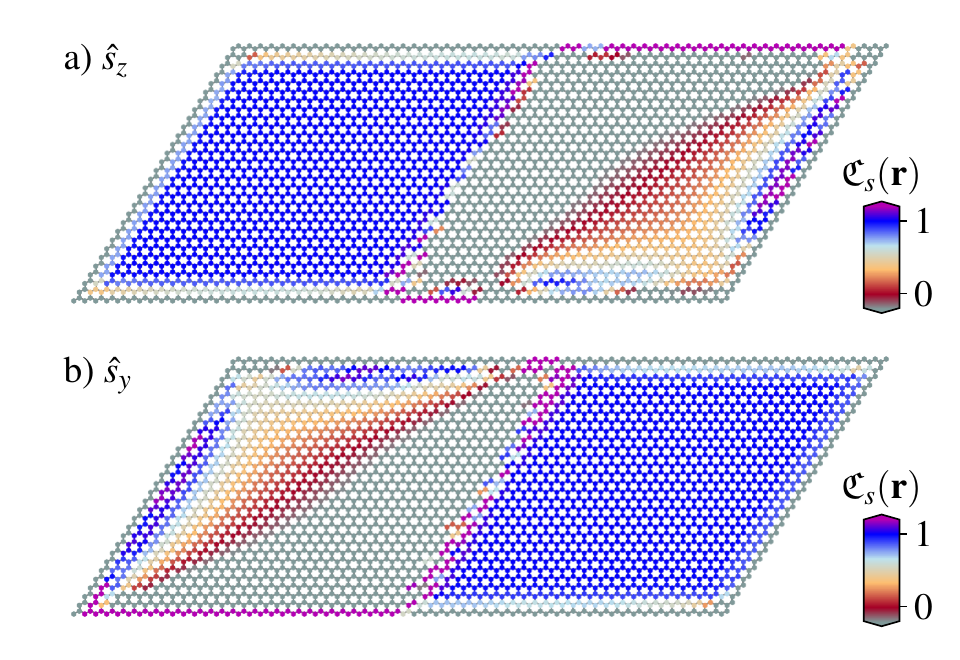}
    \caption{LSCM calculation in an inhomogeneous finite system with OBC, composed of two parts (a heterojunction): The first part (left half of the system) is described by lattice Hamiltonian $\mathcal{H}_{\rm Gen1}$ with parameters $(V_{\rm AB}/t,\ \lambda_{\rm R}/t,\ \lambda^{\rm A}_{\rm KM}/t,\ \lambda^{\rm B}_{\rm KM}/t)=(0.0,\ 0.35,\ 0.19,\ 0.21)$, while the second part (right half of the system) is described by lattice Hamiltonian $\mathcal{H}_{\rm Gen2}$ with parameters $(\lambda_{\rm R}/t,\ \lambda^{\rm A}_{\rm y}/t,\ \lambda^{\rm B}_{\rm y}/t)=(0.35,\ 0.19,\ 0.21)$. In panel (a), the projector $\mathcal{P}_{\sigma}$ [Eq. (\ref{Psigma})] was constructed from states $\ket{\psi_v}$ [Eq. (\ref{Rotsigma})] rotated with respect to the valence-projected spin matrix of the $\hat{s}_z$ operator. In panel (b), an analogous calculation was performed using the valence-projected spin matrix of the $\hat{s}_y$ operator.}
    \label{fig:Lateral}
\end{figure}

Finally, we show in Fig. \ref{fig:Lateral} the LSCM for an inhomogeneous system with OBC, composed of two parts: one described by the lattice Hamiltonian $\mathcal{H}_{\rm Gen1}$ and the other by $\mathcal{H}_{\rm Gen2}$. Such a system could model graphene grown on top of a lateral heterojunction fabricated with appropriate transition-metal dichalcogenides \cite{Lateral-HeterostructureDuan2014}. We calculate the LSCM constructing $\mathcal{P}_{\sigma}$ in Eq.(\ref{Psigma}) from states $\ket{\psi_v}$ rotated with respect to both $\hat{s}_z$ (as done in Sec. \ref{SEC4}) and $\hat{s}_y$ (as done in Sec. \ref{SEC5}) valence-preojected spin matrices. When $\hat{s}_z$ operator is used to construct the projectors $\mathcal{P}_{\sigma}$, the central region of the left half of the system, governed by $\mathcal{H}_{\rm Gen1}$, has a non-trivial quantized LSCM ($\mathfrak{C}_s\approx 1$), while the right part of the system, governed by $\mathcal{H}_{\rm Gen2}$, present an erratic behavior of the LSCM [Fig. \ref{fig:Lateral} (a)]. This erratic behavior arises from the closing of the gap $\Delta^{s_z}$ in the valence-projected spin matrix spectrum $\xi^{s_z}_v$ in the region governed by $\mathcal{H}_{\rm Gen2}$. It is analogous to the erratic behavior shown in Fig. \ref{fig:inplane-PD} (b) for $\lambda_{\rm R}>1.6 \sqrt{3}$.
The opposite occurs when $\hat{s}_y$-operator is used to construct the projectors in Eq. (\ref{Psigma}); in this case, the central region of the part of flake governed by $\mathcal{H}_{\rm Gen2}$ has a non-trivial LSCM ($\mathfrak{C}_s\approx 1$), and the one governed by $\mathcal{H}_{\rm Gen1}$ has an erratic behavior of for the marker [Fig. \ref{fig:Lateral} (b)] due to the closing of gap $\Delta^{s_y}$ in this part of the system. It is worth noting that the topology of inhomogeneous systems, such as the one presented in Fig. \ref{fig:Lateral}, cannot be described in ${\bf k}$-space, as the definition of a BZ is not possible.

\section{Final Remarks and Conclusions \label{SEC7}}

In summary, we have used the local spin Chern marker to study the real-space topology of electrons described by three distinct lattice Hamiltonians in a honeycomb system. The expression for the local spin Chern marker used here is equivalent to the one recently introduced in Ref. \cite{Bau-Marrazzo-SpinChernMarker} and is closely related to those proposed in other independent works \cite{Chen_2023, Wieder-Bradlyn-NatCommun}. The Hamiltonians considered here find applications in the field of graphene van der Waals heterostructures and are characterized by strong Rashba spin-orbit interaction, which leads to the non-conservation of the spin operator. We show that the local spin Chern marker successfully describes the topological phase diagram for the three spin-orbit couplings responsible for the manifestation of quantum spin Hall insulator physics considered in this work. The phase diagrams obtained are consistent with those calculated using the ${\bf k}$-space formula for the $\mathbb{Z}_2$-invariant. Additionally, we calculate the energy spectrum and the valence-projected spin spectrum, demonstrating that in most of the parameter space, both exhibit gaps that protect the marker against smooth variations in the model Hamiltonian.

Among the results presented in this work, we highlight those shown in Figs. \ref{fig:in-plane-LSCM-Gaps} and \ref{fig:inplane-PD}. As mentioned in the text, the robustness of the gaps $\Delta^{\rm E}$ and $\Delta^{s_y}$ produced by in-plane Kane-Mele coupling [Eq. (\ref{syKMcouplingSubLat})] against the presence of the Rashba effect strengthens the possibility of observing the quantum spin Hall insulator phase in experiments. In this situation, the quantum spin-Hall insulator phase is characterized by an in-plane polarized spin current produced by this unconventional type of spin-orbit coupling. 

It is also instructive to digress on the connection between local markers and measurable physical quantities. This connection is often subtle and depends on some mathematical properties in the marker’s definition. For instance, Ref. \cite{Resta-Bianco-PhysRevLett.110.087202} proposed an expression for macroscopic orbital magnetization \cite{Marrazzo2016-PRL116,Bianco2016-PRB93,Drigo2020-PRB101} as the trace of a local marker that plays the role of the magnetic dipole density. However, the cyclic invariance of the trace operation over matrix products allows for different expressions of this marker \cite{Seleznev-Vanderbilt-PhysRevB.107.115102}, creating ambiguity in its definition and reflecting the fact that microscopic magnetic dipole densities are not physically well-defined. In this case, only the macroscopic average of the marker has physical significance, as it is connected to orbital magnetization \cite{Resta-Bianco-PhysRevLett.110.087202} and, more recently, to hinge-bound currents \cite{Seleznev-Vanderbilt-PhysRevB.107.115102}. Such ambiguity is not present in the case of the usual local Chern marker, which can be related to the geometric part of local anomalous Hall conductivity \cite{Marrazzo2017-PRB,AHC-PhysRevB.98.115108, Seleznev-Vanderbilt-PhysRevB.107.115102}. An in-depth analysis of the mathematical properties of the local marker is beyond the scope of this paper and will be left for future work, though we offer some brief comments. A local spin Hall conductivity can indeed be defined within the framework of linear response theory \cite{Mertig-PhysRevB.101.064206}. However, even if a connection could be established, direct proportionality between the local spin Chern marker and the local spin Hall conductivity is not expected due to the presence of Rashba spin-orbit coupling. Indeed, this already occurs in the thermodynamic limit, where one can apply the band theory paradigm (${\bf k}$-space formulation). In general, finite Rashba coupling in the quantum spin Hall insulating phase causes the spin Hall conductivity to lose its quantization; however, as long as it does not close energy or spin gaps, the topological properties captured by the spin Chern numbers are not destroyed \cite{Kane-Mele-PhysRevLett.95.226801, Kane-Mele-PhysRevLett.95.146802}.

To conclude, our work broadens the scope of spin Chern numbers \cite{Prodan-PhysRevB.80.125327} as an index that accurately captures the topology associated with the quantum spin Hall insulator phase. This is achieved using the real-space version of the index, demonstrating the flexibility of this concept. The results presented here may be useful for future studies on the real-space characterization of the topology of complex materials.

\section*{ACKNOWLEDGMENTS}
The authors are grateful to the Brazilian Agencies Conselho Nacional de Desenvolvimento Cient\'\i fico e Tecnol\'ogico (CNPq), Coordena\c c\~ao de Aperfei\c coamento de Pessoal de Ensino Superior (CAPES).


\bibliography{ref}

\begin{thebibliography}{67}%
\makeatletter
\providecommand \@ifxundefined [1]{%
 \@ifx{#1\undefined}
}%
\providecommand \@ifnum [1]{%
 \ifnum #1\expandafter \@firstoftwo
 \else \expandafter \@secondoftwo
 \fi
}%
\providecommand \@ifx [1]{%
 \ifx #1\expandafter \@firstoftwo
 \else \expandafter \@secondoftwo
 \fi
}%
\providecommand \natexlab [1]{#1}%
\providecommand \enquote  [1]{``#1''}%
\providecommand \bibnamefont  [1]{#1}%
\providecommand \bibfnamefont [1]{#1}%
\providecommand \citenamefont [1]{#1}%
\providecommand \href@noop [0]{\@secondoftwo}%
\providecommand \href [0]{\begingroup \@sanitize@url \@href}%
\providecommand \@href[1]{\@@startlink{#1}\@@href}%
\providecommand \@@href[1]{\endgroup#1\@@endlink}%
\providecommand \@sanitize@url [0]{\catcode `\\12\catcode `\$12\catcode `\&12\catcode `\#12\catcode `\^12\catcode `\_12\catcode `\%12\relax}%
\providecommand \@@startlink[1]{}%
\providecommand \@@endlink[0]{}%
\providecommand \url  [0]{\begingroup\@sanitize@url \@url }%
\providecommand \@url [1]{\endgroup\@href {#1}{\urlprefix }}%
\providecommand \urlprefix  [0]{URL }%
\providecommand \Eprint [0]{\href }%
\providecommand \doibase [0]{https://doi.org/}%
\providecommand \selectlanguage [0]{\@gobble}%
\providecommand \bibinfo  [0]{\@secondoftwo}%
\providecommand \bibfield  [0]{\@secondoftwo}%
\providecommand \translation [1]{[#1]}%
\providecommand \BibitemOpen [0]{}%
\providecommand \bibitemStop [0]{}%
\providecommand \bibitemNoStop [0]{.\EOS\space}%
\providecommand \EOS [0]{\spacefactor3000\relax}%
\providecommand \BibitemShut  [1]{\csname bibitem#1\endcsname}%
\let\auto@bib@innerbib\@empty
\bibitem [{\citenamefont {Hasan}\ and\ \citenamefont {Kane}(2010)}]{Hasan-Kane-RevModPhys.82.3045}%
  \BibitemOpen
  \bibfield  {author} {\bibinfo {author} {\bibfnamefont {M.~Z.}\ \bibnamefont {Hasan}}\ and\ \bibinfo {author} {\bibfnamefont {C.~L.}\ \bibnamefont {Kane}},\ }\bibfield  {title} {\bibinfo {title} {{Colloquium: Topological insulators}},\ }\href {https://doi.org/10.1103/RevModPhys.82.3045} {\bibfield  {journal} {\bibinfo  {journal} {Rev. Mod. Phys.}\ }\textbf {\bibinfo {volume} {82}},\ \bibinfo {pages} {3045} (\bibinfo {year} {2010})}\BibitemShut {NoStop}%
\bibitem [{\citenamefont {Rachel}(2018)}]{Rachel_ReportProgressInPhysics}%
  \BibitemOpen
  \bibfield  {author} {\bibinfo {author} {\bibfnamefont {S.}~\bibnamefont {Rachel}},\ }\bibfield  {title} {\bibinfo {title} {{Interacting topological insulators: a review}},\ }\href {https://doi.org/10.1088/1361-6633/aad6a6} {\bibfield  {journal} {\bibinfo  {journal} {Reports on Progress in Physics}\ }\textbf {\bibinfo {volume} {81}},\ \bibinfo {pages} {116501} (\bibinfo {year} {2018})}\BibitemShut {NoStop}%
\bibitem [{\citenamefont {Melo}\ \emph {et~al.}(2023)\citenamefont {Melo}, \citenamefont {J\'unior}, \citenamefont {Chen}, \citenamefont {Mondaini},\ and\ \citenamefont {Paiva}}]{PMelo-PhysRevB.108.195151}%
  \BibitemOpen
  \bibfield  {author} {\bibinfo {author} {\bibfnamefont {P.~B.}\ \bibnamefont {Melo}}, \bibinfo {author} {\bibfnamefont {S.~a. A.~S.}\ \bibnamefont {J\'unior}}, \bibinfo {author} {\bibfnamefont {W.}~\bibnamefont {Chen}}, \bibinfo {author} {\bibfnamefont {R.}~\bibnamefont {Mondaini}},\ and\ \bibinfo {author} {\bibfnamefont {T.}~\bibnamefont {Paiva}},\ }\bibfield  {title} {\bibinfo {title} {{Topological marker approach to an interacting Su-Schrieffer-Heeger model}},\ }\href {https://doi.org/10.1103/PhysRevB.108.195151} {\bibfield  {journal} {\bibinfo  {journal} {Phys. Rev. B}\ }\textbf {\bibinfo {volume} {108}},\ \bibinfo {pages} {195151} (\bibinfo {year} {2023})}\BibitemShut {NoStop}%
\bibitem [{\citenamefont {Yi}\ \emph {et~al.}(2021)\citenamefont {Yi}, \citenamefont {Hu}, \citenamefont {Castro},\ and\ \citenamefont {Mondaini}}]{Mondaini-PhysRevB.104.195117}%
  \BibitemOpen
  \bibfield  {author} {\bibinfo {author} {\bibfnamefont {T.-C.}\ \bibnamefont {Yi}}, \bibinfo {author} {\bibfnamefont {S.}~\bibnamefont {Hu}}, \bibinfo {author} {\bibfnamefont {E.~V.}\ \bibnamefont {Castro}},\ and\ \bibinfo {author} {\bibfnamefont {R.}~\bibnamefont {Mondaini}},\ }\bibfield  {title} {\bibinfo {title} {{Interplay of interactions, disorder, and topology in the Haldane-Hubbard model}},\ }\href {https://doi.org/10.1103/PhysRevB.104.195117} {\bibfield  {journal} {\bibinfo  {journal} {Phys. Rev. B}\ }\textbf {\bibinfo {volume} {104}},\ \bibinfo {pages} {195117} (\bibinfo {year} {2021})}\BibitemShut {NoStop}%
\bibitem [{\citenamefont {Gilardoni}\ \emph {et~al.}(2022)\citenamefont {Gilardoni}, \citenamefont {Becca}, \citenamefont {Marrazzo},\ and\ \citenamefont {Parola}}]{Gilardoni-PhysRevB.106.L161106}%
  \BibitemOpen
  \bibfield  {author} {\bibinfo {author} {\bibfnamefont {I.}~\bibnamefont {Gilardoni}}, \bibinfo {author} {\bibfnamefont {F.}~\bibnamefont {Becca}}, \bibinfo {author} {\bibfnamefont {A.}~\bibnamefont {Marrazzo}},\ and\ \bibinfo {author} {\bibfnamefont {A.}~\bibnamefont {Parola}},\ }\bibfield  {title} {\bibinfo {title} {{Real-space many-body marker for correlated ${\mathbb{Z}}_{2}$ topological insulators}},\ }\href {https://doi.org/10.1103/PhysRevB.106.L161106} {\bibfield  {journal} {\bibinfo  {journal} {Phys. Rev. B}\ }\textbf {\bibinfo {volume} {106}},\ \bibinfo {pages} {L161106} (\bibinfo {year} {2022})}\BibitemShut {NoStop}%
\bibitem [{\citenamefont {Haldane}(1988)}]{Haldane-PhysRevLett.61.2015}%
  \BibitemOpen
  \bibfield  {author} {\bibinfo {author} {\bibfnamefont {F.~D.~M.}\ \bibnamefont {Haldane}},\ }\bibfield  {title} {\bibinfo {title} {{Model for a Quantum Hall Effect without Landau Levels: Condensed-Matter Realization of the "Parity Anomaly"}},\ }\href {https://doi.org/10.1103/PhysRevLett.61.2015} {\bibfield  {journal} {\bibinfo  {journal} {Phys. Rev. Lett.}\ }\textbf {\bibinfo {volume} {61}},\ \bibinfo {pages} {2015} (\bibinfo {year} {1988})}\BibitemShut {NoStop}%
\bibitem [{\citenamefont {Thouless}\ \emph {et~al.}(1982)\citenamefont {Thouless}, \citenamefont {Kohmoto}, \citenamefont {Nightingale},\ and\ \citenamefont {den Nijs}}]{Thouless-PhysRevLett.49.405}%
  \BibitemOpen
  \bibfield  {author} {\bibinfo {author} {\bibfnamefont {D.~J.}\ \bibnamefont {Thouless}}, \bibinfo {author} {\bibfnamefont {M.}~\bibnamefont {Kohmoto}}, \bibinfo {author} {\bibfnamefont {M.~P.}\ \bibnamefont {Nightingale}},\ and\ \bibinfo {author} {\bibfnamefont {M.}~\bibnamefont {den Nijs}},\ }\bibfield  {title} {\bibinfo {title} {{Quantized Hall Conductance in a Two-Dimensional Periodic Potential}},\ }\href {https://doi.org/10.1103/PhysRevLett.49.405} {\bibfield  {journal} {\bibinfo  {journal} {Phys. Rev. Lett.}\ }\textbf {\bibinfo {volume} {49}},\ \bibinfo {pages} {405} (\bibinfo {year} {1982})}\BibitemShut {NoStop}%
\bibitem [{\citenamefont {Kane}\ and\ \citenamefont {Mele}(2005{\natexlab{a}})}]{Kane-Mele-PhysRevLett.95.226801}%
  \BibitemOpen
  \bibfield  {author} {\bibinfo {author} {\bibfnamefont {C.~L.}\ \bibnamefont {Kane}}\ and\ \bibinfo {author} {\bibfnamefont {E.~J.}\ \bibnamefont {Mele}},\ }\bibfield  {title} {\bibinfo {title} {{Quantum Spin Hall Effect in Graphene}},\ }\href {https://doi.org/10.1103/PhysRevLett.95.226801} {\bibfield  {journal} {\bibinfo  {journal} {Phys. Rev. Lett.}\ }\textbf {\bibinfo {volume} {95}},\ \bibinfo {pages} {226801} (\bibinfo {year} {2005}{\natexlab{a}})}\BibitemShut {NoStop}%
\bibitem [{\citenamefont {Kane}\ and\ \citenamefont {Mele}(2005{\natexlab{b}})}]{Kane-Mele-PhysRevLett.95.146802}%
  \BibitemOpen
  \bibfield  {author} {\bibinfo {author} {\bibfnamefont {C.~L.}\ \bibnamefont {Kane}}\ and\ \bibinfo {author} {\bibfnamefont {E.~J.}\ \bibnamefont {Mele}},\ }\bibfield  {title} {\bibinfo {title} {{${Z}_{2}$ Topological Order and the Quantum Spin Hall Effect}},\ }\href {https://doi.org/10.1103/PhysRevLett.95.146802} {\bibfield  {journal} {\bibinfo  {journal} {Phys. Rev. Lett.}\ }\textbf {\bibinfo {volume} {95}},\ \bibinfo {pages} {146802} (\bibinfo {year} {2005}{\natexlab{b}})}\BibitemShut {NoStop}%
\bibitem [{\citenamefont {Prodan}(2009)}]{Prodan-PhysRevB.80.125327}%
  \BibitemOpen
  \bibfield  {author} {\bibinfo {author} {\bibfnamefont {E.}~\bibnamefont {Prodan}},\ }\bibfield  {title} {\bibinfo {title} {{Robustness of the spin-Chern number}},\ }\href {https://doi.org/10.1103/PhysRevB.80.125327} {\bibfield  {journal} {\bibinfo  {journal} {Phys. Rev. B}\ }\textbf {\bibinfo {volume} {80}},\ \bibinfo {pages} {125327} (\bibinfo {year} {2009})}\BibitemShut {NoStop}%
\bibitem [{\citenamefont {Soluyanov}\ and\ \citenamefont {Vanderbilt}(2012)}]{Vanderbilt-PhysRevB.85.115415}%
  \BibitemOpen
  \bibfield  {author} {\bibinfo {author} {\bibfnamefont {A.~A.}\ \bibnamefont {Soluyanov}}\ and\ \bibinfo {author} {\bibfnamefont {D.}~\bibnamefont {Vanderbilt}},\ }\bibfield  {title} {\bibinfo {title} {{Smooth gauge for topological insulators}},\ }\href {https://doi.org/10.1103/PhysRevB.85.115415} {\bibfield  {journal} {\bibinfo  {journal} {Phys. Rev. B}\ }\textbf {\bibinfo {volume} {85}},\ \bibinfo {pages} {115415} (\bibinfo {year} {2012})}\BibitemShut {NoStop}%
\bibitem [{\citenamefont {Li}\ \emph {et~al.}(2010)\citenamefont {Li}, \citenamefont {Sheng}, \citenamefont {Sheng},\ and\ \citenamefont {Xing}}]{Sheng-PhysRevB.82.165104}%
  \BibitemOpen
  \bibfield  {author} {\bibinfo {author} {\bibfnamefont {H.}~\bibnamefont {Li}}, \bibinfo {author} {\bibfnamefont {L.}~\bibnamefont {Sheng}}, \bibinfo {author} {\bibfnamefont {D.~N.}\ \bibnamefont {Sheng}},\ and\ \bibinfo {author} {\bibfnamefont {D.~Y.}\ \bibnamefont {Xing}},\ }\bibfield  {title} {\bibinfo {title} {{Chern number of thin films of the topological insulator ${\text{Bi}}_{2}{\text{Se}}_{3}$}},\ }\href {https://doi.org/10.1103/PhysRevB.82.165104} {\bibfield  {journal} {\bibinfo  {journal} {Phys. Rev. B}\ }\textbf {\bibinfo {volume} {82}},\ \bibinfo {pages} {165104} (\bibinfo {year} {2010})}\BibitemShut {NoStop}%
\bibitem [{\citenamefont {Cysne}\ \emph {et~al.}(2021)\citenamefont {Cysne}, \citenamefont {Costa}, \citenamefont {Canonico}, \citenamefont {Nardelli}, \citenamefont {Muniz},\ and\ \citenamefont {Rappoport}}]{Cysne-PhysRevLett.126.056601}%
  \BibitemOpen
  \bibfield  {author} {\bibinfo {author} {\bibfnamefont {T.~P.}\ \bibnamefont {Cysne}}, \bibinfo {author} {\bibfnamefont {M.}~\bibnamefont {Costa}}, \bibinfo {author} {\bibfnamefont {L.~M.}\ \bibnamefont {Canonico}}, \bibinfo {author} {\bibfnamefont {M.~B.}\ \bibnamefont {Nardelli}}, \bibinfo {author} {\bibfnamefont {R.~B.}\ \bibnamefont {Muniz}},\ and\ \bibinfo {author} {\bibfnamefont {T.~G.}\ \bibnamefont {Rappoport}},\ }\bibfield  {title} {\bibinfo {title} {{Disentangling Orbital and Valley Hall Effects in Bilayers of Transition Metal Dichalcogenides}},\ }\href {https://doi.org/10.1103/PhysRevLett.126.056601} {\bibfield  {journal} {\bibinfo  {journal} {Phys. Rev. Lett.}\ }\textbf {\bibinfo {volume} {126}},\ \bibinfo {pages} {056601} (\bibinfo {year} {2021})}\BibitemShut {NoStop}%
\bibitem [{\citenamefont {Cysne}\ \emph {et~al.}(2022)\citenamefont {Cysne}, \citenamefont {Bhowal}, \citenamefont {Vignale},\ and\ \citenamefont {Rappoport}}]{Cysne-PhysRevB.105.195421}%
  \BibitemOpen
  \bibfield  {author} {\bibinfo {author} {\bibfnamefont {T.~P.}\ \bibnamefont {Cysne}}, \bibinfo {author} {\bibfnamefont {S.}~\bibnamefont {Bhowal}}, \bibinfo {author} {\bibfnamefont {G.}~\bibnamefont {Vignale}},\ and\ \bibinfo {author} {\bibfnamefont {T.~G.}\ \bibnamefont {Rappoport}},\ }\bibfield  {title} {\bibinfo {title} {{Orbital Hall effect in bilayer transition metal dichalcogenides: From the intra-atomic approximation to the Bloch states orbital magnetic moment approach}},\ }\href {https://doi.org/10.1103/PhysRevB.105.195421} {\bibfield  {journal} {\bibinfo  {journal} {Phys. Rev. B}\ }\textbf {\bibinfo {volume} {105}},\ \bibinfo {pages} {195421} (\bibinfo {year} {2022})}\BibitemShut {NoStop}%
\bibitem [{\citenamefont {Lima}\ and\ \citenamefont {Lewenkopf}(2022)}]{Caio-PhysRevB.106.245408}%
  \BibitemOpen
  \bibfield  {author} {\bibinfo {author} {\bibfnamefont {L.~R.~F.}\ \bibnamefont {Lima}}\ and\ \bibinfo {author} {\bibfnamefont {C.}~\bibnamefont {Lewenkopf}},\ }\bibfield  {title} {\bibinfo {title} {{Breakdown of topological protection due to nonmagnetic edge disorder in two-dimensional materials in the quantum spin Hall phase}},\ }\href {https://doi.org/10.1103/PhysRevB.106.245408} {\bibfield  {journal} {\bibinfo  {journal} {Phys. Rev. B}\ }\textbf {\bibinfo {volume} {106}},\ \bibinfo {pages} {245408} (\bibinfo {year} {2022})}\BibitemShut {NoStop}%
\bibitem [{\citenamefont {Fonseca}\ \emph {et~al.}(2024)\citenamefont {Fonseca}, \citenamefont {Barbosa},\ and\ \citenamefont {Pereira}}]{Felipe-PhysRevB.110.075421}%
  \BibitemOpen
  \bibfield  {author} {\bibinfo {author} {\bibfnamefont {D.~B.}\ \bibnamefont {Fonseca}}, \bibinfo {author} {\bibfnamefont {A.~L.~R.}\ \bibnamefont {Barbosa}},\ and\ \bibinfo {author} {\bibfnamefont {L.~F.~C.}\ \bibnamefont {Pereira}},\ }\bibfield  {title} {\bibinfo {title} {{L\'evy flight for electrons in graphene in the presence of regions with enhanced spin-orbit coupling}},\ }\href {https://doi.org/10.1103/PhysRevB.110.075421} {\bibfield  {journal} {\bibinfo  {journal} {Phys. Rev. B}\ }\textbf {\bibinfo {volume} {110}},\ \bibinfo {pages} {075421} (\bibinfo {year} {2024})}\BibitemShut {NoStop}%
\bibitem [{\citenamefont {Fonseca}\ \emph {et~al.}(2023)\citenamefont {Fonseca}, \citenamefont {Pereira},\ and\ \citenamefont {Barbosa}}]{Anderson-PhysRevB.108.245105}%
  \BibitemOpen
  \bibfield  {author} {\bibinfo {author} {\bibfnamefont {D.~B.}\ \bibnamefont {Fonseca}}, \bibinfo {author} {\bibfnamefont {L.~L.~A.}\ \bibnamefont {Pereira}},\ and\ \bibinfo {author} {\bibfnamefont {A.~L.~R.}\ \bibnamefont {Barbosa}},\ }\bibfield  {title} {\bibinfo {title} {{Orbital Hall effect in mesoscopic devices}},\ }\href {https://doi.org/10.1103/PhysRevB.108.245105} {\bibfield  {journal} {\bibinfo  {journal} {Phys. Rev. B}\ }\textbf {\bibinfo {volume} {108}},\ \bibinfo {pages} {245105} (\bibinfo {year} {2023})}\BibitemShut {NoStop}%
\bibitem [{\citenamefont {Cerjan}\ \emph {et~al.}(2024)\citenamefont {Cerjan}, \citenamefont {Loring},\ and\ \citenamefont {Schulz-Baldes}}]{Hermann-PhysRevLett.132.073803}%
  \BibitemOpen
  \bibfield  {author} {\bibinfo {author} {\bibfnamefont {A.}~\bibnamefont {Cerjan}}, \bibinfo {author} {\bibfnamefont {T.~A.}\ \bibnamefont {Loring}},\ and\ \bibinfo {author} {\bibfnamefont {H.}~\bibnamefont {Schulz-Baldes}},\ }\bibfield  {title} {\bibinfo {title} {{Local Markers for Crystalline Topology}},\ }\href {https://doi.org/10.1103/PhysRevLett.132.073803} {\bibfield  {journal} {\bibinfo  {journal} {Phys. Rev. Lett.}\ }\textbf {\bibinfo {volume} {132}},\ \bibinfo {pages} {073803} (\bibinfo {year} {2024})}\BibitemShut {NoStop}%
\bibitem [{\citenamefont {Li}\ and\ \citenamefont {Mong}(2019)}]{Mong-PhysRevB.100.205101}%
  \BibitemOpen
  \bibfield  {author} {\bibinfo {author} {\bibfnamefont {Z.}~\bibnamefont {Li}}\ and\ \bibinfo {author} {\bibfnamefont {R.~S.~K.}\ \bibnamefont {Mong}},\ }\bibfield  {title} {\bibinfo {title} {{Local formula for the ${\mathbb{Z}}_{2}$ invariant of topological insulators}},\ }\href {https://doi.org/10.1103/PhysRevB.100.205101} {\bibfield  {journal} {\bibinfo  {journal} {Phys. Rev. B}\ }\textbf {\bibinfo {volume} {100}},\ \bibinfo {pages} {205101} (\bibinfo {year} {2019})}\BibitemShut {NoStop}%
\bibitem [{\citenamefont {Varjas}\ \emph {et~al.}(2020)\citenamefont {Varjas}, \citenamefont {Fruchart}, \citenamefont {Akhmerov},\ and\ \citenamefont {Perez-Piskunow}}]{Piskunow-PhysRevResearch.2.013229}%
  \BibitemOpen
  \bibfield  {author} {\bibinfo {author} {\bibfnamefont {D.}~\bibnamefont {Varjas}}, \bibinfo {author} {\bibfnamefont {M.}~\bibnamefont {Fruchart}}, \bibinfo {author} {\bibfnamefont {A.~R.}\ \bibnamefont {Akhmerov}},\ and\ \bibinfo {author} {\bibfnamefont {P.~M.}\ \bibnamefont {Perez-Piskunow}},\ }\bibfield  {title} {\bibinfo {title} {{Computation of topological phase diagram of disordered ${\mathrm{Pb}}_{1\ensuremath{-}x}{\mathrm{Sn}}_{x}\mathrm{Te}$ using the kernel polynomial method}},\ }\href {https://doi.org/10.1103/PhysRevResearch.2.013229} {\bibfield  {journal} {\bibinfo  {journal} {Phys. Rev. Res.}\ }\textbf {\bibinfo {volume} {2}},\ \bibinfo {pages} {013229} (\bibinfo {year} {2020})}\BibitemShut {NoStop}%
\bibitem [{\citenamefont {Kim}\ \emph {et~al.}(2023)\citenamefont {Kim}, \citenamefont {Jeon}, \citenamefont {Park},\ and\ \citenamefont {Kim}}]{HOMarker-Kim2023}%
  \BibitemOpen
  \bibfield  {author} {\bibinfo {author} {\bibfnamefont {S.-W.}\ \bibnamefont {Kim}}, \bibinfo {author} {\bibfnamefont {S.}~\bibnamefont {Jeon}}, \bibinfo {author} {\bibfnamefont {M.~J.}\ \bibnamefont {Park}},\ and\ \bibinfo {author} {\bibfnamefont {Y.}~\bibnamefont {Kim}},\ }\bibfield  {title} {\bibinfo {title} {{Replica higher-order topology of Hofstadter butterflies in twisted bilayer graphene}},\ }\bibfield  {journal} {\bibinfo  {journal} {npj Computational Materials}\ }\textbf {\bibinfo {volume} {9}},\ \href {https://doi.org/10.1038/s41524-023-01105-5} {10.1038/s41524-023-01105-5} (\bibinfo {year} {2023})\BibitemShut {NoStop}%
\bibitem [{\citenamefont {Regis}\ \emph {et~al.}(2024)\citenamefont {Regis}, \citenamefont {Velasco}, \citenamefont {Neto},\ and\ \citenamefont {Lewenkopf}}]{Regis2024}%
  \BibitemOpen
  \bibfield  {author} {\bibinfo {author} {\bibfnamefont {V.}~\bibnamefont {Regis}}, \bibinfo {author} {\bibfnamefont {V.}~\bibnamefont {Velasco}}, \bibinfo {author} {\bibfnamefont {M.~B.~S.}\ \bibnamefont {Neto}},\ and\ \bibinfo {author} {\bibfnamefont {C.}~\bibnamefont {Lewenkopf}},\ }\bibfield  {title} {\bibinfo {title} {Structure-driven phase transitions in paracrystalline topological insulators},\ }\href {https://doi.org/10.1103/PhysRevB.110.L161105} {\bibfield  {journal} {\bibinfo  {journal} {Phys. Rev. B}\ }\textbf {\bibinfo {volume} {110}},\ \bibinfo {pages} {L161105} (\bibinfo {year} {2024})}\BibitemShut {NoStop}%
\bibitem [{\citenamefont {Bianco}\ and\ \citenamefont {Resta}(2011)}]{Bianco-Resta-PhysRevB.84.241106}%
  \BibitemOpen
  \bibfield  {author} {\bibinfo {author} {\bibfnamefont {R.}~\bibnamefont {Bianco}}\ and\ \bibinfo {author} {\bibfnamefont {R.}~\bibnamefont {Resta}},\ }\bibfield  {title} {\bibinfo {title} {{Mapping topological order in coordinate space}},\ }\href {https://doi.org/10.1103/PhysRevB.84.241106} {\bibfield  {journal} {\bibinfo  {journal} {Phys. Rev. B}\ }\textbf {\bibinfo {volume} {84}},\ \bibinfo {pages} {241106} (\bibinfo {year} {2011})}\BibitemShut {NoStop}%
\bibitem [{\citenamefont {Ba\`u}\ and\ \citenamefont {Marrazzo}(2024{\natexlab{a}})}]{Bau-Marrazzo-PhysRevB.109.014206}%
  \BibitemOpen
  \bibfield  {author} {\bibinfo {author} {\bibfnamefont {N.}~\bibnamefont {Ba\`u}}\ and\ \bibinfo {author} {\bibfnamefont {A.}~\bibnamefont {Marrazzo}},\ }\bibfield  {title} {\bibinfo {title} {{Local Chern marker for periodic systems}},\ }\href {https://doi.org/10.1103/PhysRevB.109.014206} {\bibfield  {journal} {\bibinfo  {journal} {Phys. Rev. B}\ }\textbf {\bibinfo {volume} {109}},\ \bibinfo {pages} {014206} (\bibinfo {year} {2024}{\natexlab{a}})}\BibitemShut {NoStop}%
\bibitem [{\citenamefont {Molignini}\ \emph {et~al.}(2023)\citenamefont {Molignini}, \citenamefont {Lapierre}, \citenamefont {Chitra},\ and\ \citenamefont {Chen}}]{Molignini2023-TempChernMarker}%
  \BibitemOpen
  \bibfield  {author} {\bibinfo {author} {\bibfnamefont {P.}~\bibnamefont {Molignini}}, \bibinfo {author} {\bibfnamefont {B.}~\bibnamefont {Lapierre}}, \bibinfo {author} {\bibfnamefont {R.}~\bibnamefont {Chitra}},\ and\ \bibinfo {author} {\bibfnamefont {W.}~\bibnamefont {Chen}},\ }\bibfield  {title} {\bibinfo {title} {{Probing Chern number by opacity and topological phase transition by a nonlocal Chern marker}},\ }\bibfield  {journal} {\bibinfo  {journal} {SciPost Physics Core}\ }\textbf {\bibinfo {volume} {6}},\ \href {https://doi.org/10.21468/scipostphyscore.6.3.059} {10.21468/scipostphyscore.6.3.059} (\bibinfo {year} {2023})\BibitemShut {NoStop}%
\bibitem [{\citenamefont {Oliveira}\ and\ \citenamefont {Chen}(2024)}]{Chen-PhysRevB.109.094202}%
  \BibitemOpen
  \bibfield  {author} {\bibinfo {author} {\bibfnamefont {L.~A.}\ \bibnamefont {Oliveira}}\ and\ \bibinfo {author} {\bibfnamefont {W.}~\bibnamefont {Chen}},\ }\bibfield  {title} {\bibinfo {title} {{Robustness of topological order against disorder}},\ }\href {https://doi.org/10.1103/PhysRevB.109.094202} {\bibfield  {journal} {\bibinfo  {journal} {Phys. Rev. B}\ }\textbf {\bibinfo {volume} {109}},\ \bibinfo {pages} {094202} (\bibinfo {year} {2024})}\BibitemShut {NoStop}%
\bibitem [{\citenamefont {Assun\ifmmode \mbox{\c{c}}\else~\c{c}\fi{}\ ao}\ \emph {et~al.}(2024)\citenamefont {Assun\ifmmode \mbox{\c{c}}\else~\c{c}\fi{}\ ao}, \citenamefont {Ferreira},\ and\ \citenamefont {Lewenkopf}}]{Caio-PhysRevB.109.L201102}%
  \BibitemOpen
  \bibfield  {author} {\bibinfo {author} {\bibfnamefont {B.~D.}\ \bibnamefont {Assun\ifmmode \mbox{\c{c}}\else~\c{c}\fi{}\ ao}}, \bibinfo {author} {\bibfnamefont {G.~J.}\ \bibnamefont {Ferreira}},\ and\ \bibinfo {author} {\bibfnamefont {C.~H.}\ \bibnamefont {Lewenkopf}},\ }\bibfield  {title} {\bibinfo {title} {{Phase transitions and scale invariance in topological Anderson insulators}},\ }\href {https://doi.org/10.1103/PhysRevB.109.L201102} {\bibfield  {journal} {\bibinfo  {journal} {Phys. Rev. B}\ }\textbf {\bibinfo {volume} {109}},\ \bibinfo {pages} {L201102} (\bibinfo {year} {2024})}\BibitemShut {NoStop}%
\bibitem [{\citenamefont {Ba\`u}\ and\ \citenamefont {Marrazzo}(2024{\natexlab{b}})}]{Bau-Marrazzo-SpinChernMarker}%
  \BibitemOpen
  \bibfield  {author} {\bibinfo {author} {\bibfnamefont {N.}~\bibnamefont {Ba\`u}}\ and\ \bibinfo {author} {\bibfnamefont {A.}~\bibnamefont {Marrazzo}},\ }\bibfield  {title} {\bibinfo {title} {{Theory of local ${\mathbb{Z}}_{2}$ topological markers for finite and periodic two-dimensional systems}},\ }\href {https://doi.org/10.1103/PhysRevB.110.054203} {\bibfield  {journal} {\bibinfo  {journal} {Phys. Rev. B}\ }\textbf {\bibinfo {volume} {110}},\ \bibinfo {pages} {054203} (\bibinfo {year} {2024}{\natexlab{b}})}\BibitemShut {NoStop}%
\bibitem [{\citenamefont {Gmitra}\ and\ \citenamefont {Fabian}(2015)}]{Fabian-PhysRevB.92.155403}%
  \BibitemOpen
  \bibfield  {author} {\bibinfo {author} {\bibfnamefont {M.}~\bibnamefont {Gmitra}}\ and\ \bibinfo {author} {\bibfnamefont {J.}~\bibnamefont {Fabian}},\ }\bibfield  {title} {\bibinfo {title} {{Graphene on transition-metal dichalcogenides: A platform for proximity spin-orbit physics and optospintronics}},\ }\href {https://doi.org/10.1103/PhysRevB.92.155403} {\bibfield  {journal} {\bibinfo  {journal} {Phys. Rev. B}\ }\textbf {\bibinfo {volume} {92}},\ \bibinfo {pages} {155403} (\bibinfo {year} {2015})}\BibitemShut {NoStop}%
\bibitem [{\citenamefont {Frank}\ \emph {et~al.}(2018)\citenamefont {Frank}, \citenamefont {H\"ogl}, \citenamefont {Gmitra}, \citenamefont {Kochan},\ and\ \citenamefont {Fabian}}]{Fabian-PhysRevLett.120.156402}%
  \BibitemOpen
  \bibfield  {author} {\bibinfo {author} {\bibfnamefont {T.}~\bibnamefont {Frank}}, \bibinfo {author} {\bibfnamefont {P.}~\bibnamefont {H\"ogl}}, \bibinfo {author} {\bibfnamefont {M.}~\bibnamefont {Gmitra}}, \bibinfo {author} {\bibfnamefont {D.}~\bibnamefont {Kochan}},\ and\ \bibinfo {author} {\bibfnamefont {J.}~\bibnamefont {Fabian}},\ }\bibfield  {title} {\bibinfo {title} {{Protected Pseudohelical Edge States in ${\mathbb{Z}}_{2}$-Trivial Proximitized Graphene}},\ }\href {https://doi.org/10.1103/PhysRevLett.120.156402} {\bibfield  {journal} {\bibinfo  {journal} {Phys. Rev. Lett.}\ }\textbf {\bibinfo {volume} {120}},\ \bibinfo {pages} {156402} (\bibinfo {year} {2018})}\BibitemShut {NoStop}%
\bibitem [{\citenamefont {Cysne}\ \emph {et~al.}(2018{\natexlab{a}})\citenamefont {Cysne}, \citenamefont {Ferreira},\ and\ \citenamefont {Rappoport}}]{Cysne-PhysRevB.98.045407}%
  \BibitemOpen
  \bibfield  {author} {\bibinfo {author} {\bibfnamefont {T.~P.}\ \bibnamefont {Cysne}}, \bibinfo {author} {\bibfnamefont {A.}~\bibnamefont {Ferreira}},\ and\ \bibinfo {author} {\bibfnamefont {T.~G.}\ \bibnamefont {Rappoport}},\ }\bibfield  {title} {\bibinfo {title} {{Crystal-field effects in graphene with interface-induced spin-orbit coupling}},\ }\href {https://doi.org/10.1103/PhysRevB.98.045407} {\bibfield  {journal} {\bibinfo  {journal} {Phys. Rev. B}\ }\textbf {\bibinfo {volume} {98}},\ \bibinfo {pages} {045407} (\bibinfo {year} {2018}{\natexlab{a}})}\BibitemShut {NoStop}%
\bibitem [{\citenamefont {Garcia}\ \emph {et~al.}(2020)\citenamefont {Garcia}, \citenamefont {Vila}, \citenamefont {Hsu}, \citenamefont {Waintal}, \citenamefont {Pereira},\ and\ \citenamefont {Roche}}]{Jose-PhysRevLett.125.256603}%
  \BibitemOpen
  \bibfield  {author} {\bibinfo {author} {\bibfnamefont {J.~H.}\ \bibnamefont {Garcia}}, \bibinfo {author} {\bibfnamefont {M.}~\bibnamefont {Vila}}, \bibinfo {author} {\bibfnamefont {C.-H.}\ \bibnamefont {Hsu}}, \bibinfo {author} {\bibfnamefont {X.}~\bibnamefont {Waintal}}, \bibinfo {author} {\bibfnamefont {V.~M.}\ \bibnamefont {Pereira}},\ and\ \bibinfo {author} {\bibfnamefont {S.}~\bibnamefont {Roche}},\ }\bibfield  {title} {\bibinfo {title} {{Canted Persistent Spin Texture and Quantum Spin Hall Effect in ${\mathrm{WTe}}_{2}$}},\ }\href {https://doi.org/10.1103/PhysRevLett.125.256603} {\bibfield  {journal} {\bibinfo  {journal} {Phys. Rev. Lett.}\ }\textbf {\bibinfo {volume} {125}},\ \bibinfo {pages} {256603} (\bibinfo {year} {2020})}\BibitemShut {NoStop}%
\bibitem [{\citenamefont {Ji}\ \emph {et~al.}(2024)\citenamefont {Ji}, \citenamefont {Quan}, \citenamefont {Yao}, \citenamefont {Yang},\ and\ \citenamefont {Li}}]{Li-PhysRevB.109.155407}%
  \BibitemOpen
  \bibfield  {author} {\bibinfo {author} {\bibfnamefont {S.}~\bibnamefont {Ji}}, \bibinfo {author} {\bibfnamefont {C.}~\bibnamefont {Quan}}, \bibinfo {author} {\bibfnamefont {R.}~\bibnamefont {Yao}}, \bibinfo {author} {\bibfnamefont {J.}~\bibnamefont {Yang}},\ and\ \bibinfo {author} {\bibfnamefont {X.}~\bibnamefont {Li}},\ }\bibfield  {title} {\bibinfo {title} {{Reversal of orbital Hall conductivity and emergence of tunable topological quantum states in orbital Hall insulators}},\ }\href {https://doi.org/10.1103/PhysRevB.109.155407} {\bibfield  {journal} {\bibinfo  {journal} {Phys. Rev. B}\ }\textbf {\bibinfo {volume} {109}},\ \bibinfo {pages} {155407} (\bibinfo {year} {2024})}\BibitemShut {NoStop}%
\bibitem [{\citenamefont {Ji}\ \emph {et~al.}(2023)\citenamefont {Ji}, \citenamefont {Yao}, \citenamefont {Quan}, \citenamefont {Wang}, \citenamefont {Yang},\ and\ \citenamefont {Li}}]{Li-PhysRevB.108.224422}%
  \BibitemOpen
  \bibfield  {author} {\bibinfo {author} {\bibfnamefont {S.}~\bibnamefont {Ji}}, \bibinfo {author} {\bibfnamefont {R.}~\bibnamefont {Yao}}, \bibinfo {author} {\bibfnamefont {C.}~\bibnamefont {Quan}}, \bibinfo {author} {\bibfnamefont {Y.}~\bibnamefont {Wang}}, \bibinfo {author} {\bibfnamefont {J.}~\bibnamefont {Yang}},\ and\ \bibinfo {author} {\bibfnamefont {X.}~\bibnamefont {Li}},\ }\bibfield  {title} {\bibinfo {title} {{Observing topological phase transition in ferromagnetic transition metal dichalcogenides}},\ }\href {https://doi.org/10.1103/PhysRevB.108.224422} {\bibfield  {journal} {\bibinfo  {journal} {Phys. Rev. B}\ }\textbf {\bibinfo {volume} {108}},\ \bibinfo {pages} {224422} (\bibinfo {year} {2023})}\BibitemShut {NoStop}%
\bibitem [{\citenamefont {Sheng}\ \emph {et~al.}(2006)\citenamefont {Sheng}, \citenamefont {Weng}, \citenamefont {Sheng},\ and\ \citenamefont {Haldane}}]{Sheng-PhysRevLett.97.036808}%
  \BibitemOpen
  \bibfield  {author} {\bibinfo {author} {\bibfnamefont {D.~N.}\ \bibnamefont {Sheng}}, \bibinfo {author} {\bibfnamefont {Z.~Y.}\ \bibnamefont {Weng}}, \bibinfo {author} {\bibfnamefont {L.}~\bibnamefont {Sheng}},\ and\ \bibinfo {author} {\bibfnamefont {F.~D.~M.}\ \bibnamefont {Haldane}},\ }\bibfield  {title} {\bibinfo {title} {{Quantum Spin-Hall Effect and Topologically Invariant Chern Numbers}},\ }\href {https://doi.org/10.1103/PhysRevLett.97.036808} {\bibfield  {journal} {\bibinfo  {journal} {Phys. Rev. Lett.}\ }\textbf {\bibinfo {volume} {97}},\ \bibinfo {pages} {036808} (\bibinfo {year} {2006})}\BibitemShut {NoStop}%
\bibitem [{\citenamefont {Yang}\ \emph {et~al.}(2011)\citenamefont {Yang}, \citenamefont {Xu}, \citenamefont {Sheng}, \citenamefont {Wang}, \citenamefont {Xing},\ and\ \citenamefont {Sheng}}]{Yang2011-PhysRevLett.107.066602}%
  \BibitemOpen
  \bibfield  {author} {\bibinfo {author} {\bibfnamefont {Y.}~\bibnamefont {Yang}}, \bibinfo {author} {\bibfnamefont {Z.}~\bibnamefont {Xu}}, \bibinfo {author} {\bibfnamefont {L.}~\bibnamefont {Sheng}}, \bibinfo {author} {\bibfnamefont {B.}~\bibnamefont {Wang}}, \bibinfo {author} {\bibfnamefont {D.~Y.}\ \bibnamefont {Xing}},\ and\ \bibinfo {author} {\bibfnamefont {D.~N.}\ \bibnamefont {Sheng}},\ }\bibfield  {title} {\bibinfo {title} {{Time-Reversal-Symmetry-Broken Quantum Spin Hall Effect}},\ }\href {https://doi.org/10.1103/PhysRevLett.107.066602} {\bibfield  {journal} {\bibinfo  {journal} {Phys. Rev. Lett.}\ }\textbf {\bibinfo {volume} {107}},\ \bibinfo {pages} {066602} (\bibinfo {year} {2011})}\BibitemShut {NoStop}%
\bibitem [{\citenamefont {Wang}\ \emph {et~al.}(2024)\citenamefont {Wang}, \citenamefont {Zhou}, \citenamefont {Hung}, \citenamefont {Lin}, \citenamefont {Lin},\ and\ \citenamefont {Bansil}}]{Baokai-2DMat-Antimonene}%
  \BibitemOpen
  \bibfield  {author} {\bibinfo {author} {\bibfnamefont {B.}~\bibnamefont {Wang}}, \bibinfo {author} {\bibfnamefont {X.}~\bibnamefont {Zhou}}, \bibinfo {author} {\bibfnamefont {Y.-C.}\ \bibnamefont {Hung}}, \bibinfo {author} {\bibfnamefont {Y.-C.}\ \bibnamefont {Lin}}, \bibinfo {author} {\bibfnamefont {H.}~\bibnamefont {Lin}},\ and\ \bibinfo {author} {\bibfnamefont {A.}~\bibnamefont {Bansil}},\ }\bibfield  {title} {\bibinfo {title} {{High spin-Chern-number insulator in $\alpha$-antimonene with a hidden topological phase}},\ }\href {https://doi.org/10.1088/2053-1583/ad3136} {\bibfield  {journal} {\bibinfo  {journal} {2D Materials}\ }\textbf {\bibinfo {volume} {11}},\ \bibinfo {pages} {025033} (\bibinfo {year} {2024})}\BibitemShut {NoStop}%
\bibitem [{\citenamefont {Lai}\ \emph {et~al.}(2024)\citenamefont {Lai}, \citenamefont {Liu}, \citenamefont {Xie}, \citenamefont {Deng}, \citenamefont {Wang}, \citenamefont {Cheng}, \citenamefont {Liu},\ and\ \citenamefont {Chen}}]{Lai-PhysRevB.109.L140104}%
  \BibitemOpen
  \bibfield  {author} {\bibinfo {author} {\bibfnamefont {P.}~\bibnamefont {Lai}}, \bibinfo {author} {\bibfnamefont {H.}~\bibnamefont {Liu}}, \bibinfo {author} {\bibfnamefont {B.}~\bibnamefont {Xie}}, \bibinfo {author} {\bibfnamefont {W.}~\bibnamefont {Deng}}, \bibinfo {author} {\bibfnamefont {H.}~\bibnamefont {Wang}}, \bibinfo {author} {\bibfnamefont {H.}~\bibnamefont {Cheng}}, \bibinfo {author} {\bibfnamefont {Z.}~\bibnamefont {Liu}},\ and\ \bibinfo {author} {\bibfnamefont {S.}~\bibnamefont {Chen}},\ }\bibfield  {title} {\bibinfo {title} {{Spin Chern insulator in a phononic fractal lattice}},\ }\href {https://doi.org/10.1103/PhysRevB.109.L140104} {\bibfield  {journal} {\bibinfo  {journal} {Phys. Rev. B}\ }\textbf {\bibinfo {volume} {109}},\ \bibinfo {pages} {L140104} (\bibinfo {year} {2024})}\BibitemShut {NoStop}%
\bibitem [{\citenamefont {Wang}\ \emph {et~al.}(2023)\citenamefont {Wang}, \citenamefont {Hung}, \citenamefont {Zhou}, \citenamefont {Bansil},\ and\ \citenamefont {Lin}}]{Wang-PhysRevB.108.245103}%
  \BibitemOpen
  \bibfield  {author} {\bibinfo {author} {\bibfnamefont {B.}~\bibnamefont {Wang}}, \bibinfo {author} {\bibfnamefont {Y.-C.}\ \bibnamefont {Hung}}, \bibinfo {author} {\bibfnamefont {X.}~\bibnamefont {Zhou}}, \bibinfo {author} {\bibfnamefont {A.}~\bibnamefont {Bansil}},\ and\ \bibinfo {author} {\bibfnamefont {H.}~\bibnamefont {Lin}},\ }\bibfield  {title} {\bibinfo {title} {{Higher-order topological phases hidden in quantum spin Hall insulators}},\ }\href {https://doi.org/10.1103/PhysRevB.108.245103} {\bibfield  {journal} {\bibinfo  {journal} {Phys. Rev. B}\ }\textbf {\bibinfo {volume} {108}},\ \bibinfo {pages} {245103} (\bibinfo {year} {2023})}\BibitemShut {NoStop}%
\bibitem [{\citenamefont {Deng}\ \emph {et~al.}(2020)\citenamefont {Deng}, \citenamefont {Huang}, \citenamefont {Lu}, \citenamefont {Peri}, \citenamefont {Li}, \citenamefont {Huber},\ and\ \citenamefont {Liu}}]{NatCommn-Deng-2020}%
  \BibitemOpen
  \bibfield  {author} {\bibinfo {author} {\bibfnamefont {W.}~\bibnamefont {Deng}}, \bibinfo {author} {\bibfnamefont {X.}~\bibnamefont {Huang}}, \bibinfo {author} {\bibfnamefont {J.}~\bibnamefont {Lu}}, \bibinfo {author} {\bibfnamefont {V.}~\bibnamefont {Peri}}, \bibinfo {author} {\bibfnamefont {F.}~\bibnamefont {Li}}, \bibinfo {author} {\bibfnamefont {S.~D.}\ \bibnamefont {Huber}},\ and\ \bibinfo {author} {\bibfnamefont {Z.}~\bibnamefont {Liu}},\ }\bibfield  {title} {\bibinfo {title} {{Acoustic spin-Chern insulator induced by synthetic spin–orbit coupling with spin conservation breaking}},\ }\bibfield  {journal} {\bibinfo  {journal} {Nature Communications}\ }\textbf {\bibinfo {volume} {11}},\ \href {https://doi.org/10.1038/s41467-020-17039-1} {10.1038/s41467-020-17039-1} (\bibinfo {year} {2020})\BibitemShut {NoStop}%
\bibitem [{\citenamefont {Wang}\ \emph {et~al.}(2020)\citenamefont {Wang}, \citenamefont {Wang}, \citenamefont {Yang}, \citenamefont {Shi}, \citenamefont {Ruan}, \citenamefont {Xing}, \citenamefont {Wang},\ and\ \citenamefont {Zhang}}]{Wang-PhysRevB.101.081109}%
  \BibitemOpen
  \bibfield  {author} {\bibinfo {author} {\bibfnamefont {H.}~\bibnamefont {Wang}}, \bibinfo {author} {\bibfnamefont {D.}~\bibnamefont {Wang}}, \bibinfo {author} {\bibfnamefont {Z.}~\bibnamefont {Yang}}, \bibinfo {author} {\bibfnamefont {M.}~\bibnamefont {Shi}}, \bibinfo {author} {\bibfnamefont {J.}~\bibnamefont {Ruan}}, \bibinfo {author} {\bibfnamefont {D.}~\bibnamefont {Xing}}, \bibinfo {author} {\bibfnamefont {J.}~\bibnamefont {Wang}},\ and\ \bibinfo {author} {\bibfnamefont {H.}~\bibnamefont {Zhang}},\ }\bibfield  {title} {\bibinfo {title} {{Dynamical axion state with hidden pseudospin Chern numbers in ${\mathrm{MnBi}}_{2}{\mathrm{Te}}_{4}$-based heterostructures}},\ }\href {https://doi.org/10.1103/PhysRevB.101.081109} {\bibfield  {journal} {\bibinfo  {journal} {Phys. Rev. B}\ }\textbf {\bibinfo {volume} {101}},\ \bibinfo {pages} {081109} (\bibinfo {year} {2020})}\BibitemShut {NoStop}%
\bibitem [{\citenamefont {Chen}(2023{\natexlab{a}})}]{Chen_2023}%
  \BibitemOpen
  \bibfield  {author} {\bibinfo {author} {\bibfnamefont {W.}~\bibnamefont {Chen}},\ }\bibfield  {title} {\bibinfo {title} {{Optical absorption measurement of spin Berry curvature and spin Chern marker}},\ }\href {https://doi.org/10.1088/1361-648X/acba72} {\bibfield  {journal} {\bibinfo  {journal} {Journal of Physics: Condensed Matter}\ }\textbf {\bibinfo {volume} {35}},\ \bibinfo {pages} {155601} (\bibinfo {year} {2023}{\natexlab{a}})}\BibitemShut {NoStop}%
\bibitem [{\citenamefont {Lin}\ \emph {et~al.}(2024)\citenamefont {Lin}, \citenamefont {Palumbo}, \citenamefont {Guo}, \citenamefont {Hwang}, \citenamefont {Blackburn}, \citenamefont {Shoemaker}, \citenamefont {Mahmood}, \citenamefont {Wang}, \citenamefont {Fiete}, \citenamefont {Wieder},\ and\ \citenamefont {Bradlyn}}]{Wieder-Bradlyn-NatCommun}%
  \BibitemOpen
  \bibfield  {author} {\bibinfo {author} {\bibfnamefont {K.-S.}\ \bibnamefont {Lin}}, \bibinfo {author} {\bibfnamefont {G.}~\bibnamefont {Palumbo}}, \bibinfo {author} {\bibfnamefont {Z.}~\bibnamefont {Guo}}, \bibinfo {author} {\bibfnamefont {Y.}~\bibnamefont {Hwang}}, \bibinfo {author} {\bibfnamefont {J.}~\bibnamefont {Blackburn}}, \bibinfo {author} {\bibfnamefont {D.~P.}\ \bibnamefont {Shoemaker}}, \bibinfo {author} {\bibfnamefont {F.}~\bibnamefont {Mahmood}}, \bibinfo {author} {\bibfnamefont {Z.}~\bibnamefont {Wang}}, \bibinfo {author} {\bibfnamefont {G.~A.}\ \bibnamefont {Fiete}}, \bibinfo {author} {\bibfnamefont {B.~J.}\ \bibnamefont {Wieder}},\ and\ \bibinfo {author} {\bibfnamefont {B.}~\bibnamefont {Bradlyn}},\ }\bibfield  {title} {\bibinfo {title} {{Spin-resolved topology and partial axion angles in three-dimensional insulators}},\ }\bibfield  {journal} {\bibinfo  {journal} {Nature Communications}\ }\textbf {\bibinfo {volume} {15}},\ \href {https://doi.org/10.1038/s41467-024-44762-w}
  {10.1038/s41467-024-44762-w} (\bibinfo {year} {2024})\BibitemShut {NoStop}%
\bibitem [{\citenamefont {Chen}(2023{\natexlab{b}})}]{Chen-PhysRevB.107.045111}%
  \BibitemOpen
  \bibfield  {author} {\bibinfo {author} {\bibfnamefont {W.}~\bibnamefont {Chen}},\ }\bibfield  {title} {\bibinfo {title} {{Universal topological marker}},\ }\href {https://doi.org/10.1103/PhysRevB.107.045111} {\bibfield  {journal} {\bibinfo  {journal} {Phys. Rev. B}\ }\textbf {\bibinfo {volume} {107}},\ \bibinfo {pages} {045111} (\bibinfo {year} {2023}{\natexlab{b}})}\BibitemShut {NoStop}%
\bibitem [{\citenamefont {Bychkov}\ and\ \citenamefont {Rashba}(1984)}]{Bychkov-Rashba_1984}%
  \BibitemOpen
  \bibfield  {author} {\bibinfo {author} {\bibfnamefont {Y.~A.}\ \bibnamefont {Bychkov}}\ and\ \bibinfo {author} {\bibfnamefont {E.~I.}\ \bibnamefont {Rashba}},\ }\bibfield  {title} {\bibinfo {title} {{Oscillatory effects and the magnetic susceptibility of carriers in inversion layers}},\ }\href {https://doi.org/10.1088/0022-3719/17/33/015} {\bibfield  {journal} {\bibinfo  {journal} {Journal of Physics C: Solid State Physics}\ }\textbf {\bibinfo {volume} {17}},\ \bibinfo {pages} {6039} (\bibinfo {year} {1984})}\BibitemShut {NoStop}%
\bibitem [{\citenamefont {Kochan}\ \emph {et~al.}(2017)\citenamefont {Kochan}, \citenamefont {Irmer},\ and\ \citenamefont {Fabian}}]{Koschan-PhysRevB.95.165415}%
  \BibitemOpen
  \bibfield  {author} {\bibinfo {author} {\bibfnamefont {D.}~\bibnamefont {Kochan}}, \bibinfo {author} {\bibfnamefont {S.}~\bibnamefont {Irmer}},\ and\ \bibinfo {author} {\bibfnamefont {J.}~\bibnamefont {Fabian}},\ }\bibfield  {title} {\bibinfo {title} {{Model spin-orbit coupling Hamiltonians for graphene systems}},\ }\href {https://doi.org/10.1103/PhysRevB.95.165415} {\bibfield  {journal} {\bibinfo  {journal} {Phys. Rev. B}\ }\textbf {\bibinfo {volume} {95}},\ \bibinfo {pages} {165415} (\bibinfo {year} {2017})}\BibitemShut {NoStop}%
\bibitem [{\citenamefont {Avsar}\ \emph {et~al.}(2020)\citenamefont {Avsar}, \citenamefont {Ochoa}, \citenamefont {Guinea}, \citenamefont {\"Ozyilmaz}, \citenamefont {van Wees},\ and\ \citenamefont {Vera-Marun}}]{RevModPhys.92.021003}%
  \BibitemOpen
  \bibfield  {author} {\bibinfo {author} {\bibfnamefont {A.}~\bibnamefont {Avsar}}, \bibinfo {author} {\bibfnamefont {H.}~\bibnamefont {Ochoa}}, \bibinfo {author} {\bibfnamefont {F.}~\bibnamefont {Guinea}}, \bibinfo {author} {\bibfnamefont {B.}~\bibnamefont {\"Ozyilmaz}}, \bibinfo {author} {\bibfnamefont {B.~J.}\ \bibnamefont {van Wees}},\ and\ \bibinfo {author} {\bibfnamefont {I.~J.}\ \bibnamefont {Vera-Marun}},\ }\bibfield  {title} {\bibinfo {title} {{Colloquium: Spintronics in graphene and other two-dimensional materials}},\ }\href {https://doi.org/10.1103/RevModPhys.92.021003} {\bibfield  {journal} {\bibinfo  {journal} {Rev. Mod. Phys.}\ }\textbf {\bibinfo {volume} {92}},\ \bibinfo {pages} {021003} (\bibinfo {year} {2020})}\BibitemShut {NoStop}%
\bibitem [{\citenamefont {Cummings}\ \emph {et~al.}(2017)\citenamefont {Cummings}, \citenamefont {Garcia}, \citenamefont {Fabian},\ and\ \citenamefont {Roche}}]{Roche-PhysRevLett.119.206601}%
  \BibitemOpen
  \bibfield  {author} {\bibinfo {author} {\bibfnamefont {A.~W.}\ \bibnamefont {Cummings}}, \bibinfo {author} {\bibfnamefont {J.~H.}\ \bibnamefont {Garcia}}, \bibinfo {author} {\bibfnamefont {J.}~\bibnamefont {Fabian}},\ and\ \bibinfo {author} {\bibfnamefont {S.}~\bibnamefont {Roche}},\ }\bibfield  {title} {\bibinfo {title} {{Giant Spin Lifetime Anisotropy in Graphene Induced by Proximity Effects}},\ }\href {https://doi.org/10.1103/PhysRevLett.119.206601} {\bibfield  {journal} {\bibinfo  {journal} {Phys. Rev. Lett.}\ }\textbf {\bibinfo {volume} {119}},\ \bibinfo {pages} {206601} (\bibinfo {year} {2017})}\BibitemShut {NoStop}%
\bibitem [{\citenamefont {Offidani}\ \emph {et~al.}(2017)\citenamefont {Offidani}, \citenamefont {Milletar\`{\i}}, \citenamefont {Raimondi},\ and\ \citenamefont {Ferreira}}]{Offidani-PhysRevLett.119.196801}%
  \BibitemOpen
  \bibfield  {author} {\bibinfo {author} {\bibfnamefont {M.}~\bibnamefont {Offidani}}, \bibinfo {author} {\bibfnamefont {M.}~\bibnamefont {Milletar\`{\i}}}, \bibinfo {author} {\bibfnamefont {R.}~\bibnamefont {Raimondi}},\ and\ \bibinfo {author} {\bibfnamefont {A.}~\bibnamefont {Ferreira}},\ }\bibfield  {title} {\bibinfo {title} {{Optimal Charge-to-Spin Conversion in Graphene on Transition-Metal Dichalcogenides}},\ }\href {https://doi.org/10.1103/PhysRevLett.119.196801} {\bibfield  {journal} {\bibinfo  {journal} {Phys. Rev. Lett.}\ }\textbf {\bibinfo {volume} {119}},\ \bibinfo {pages} {196801} (\bibinfo {year} {2017})}\BibitemShut {NoStop}%
\bibitem [{\citenamefont {Cysne}\ \emph {et~al.}(2018{\natexlab{b}})\citenamefont {Cysne}, \citenamefont {Garcia}, \citenamefont {Rocha},\ and\ \citenamefont {Rappoport}}]{Cysne-PhysRevB.97.085413}%
  \BibitemOpen
  \bibfield  {author} {\bibinfo {author} {\bibfnamefont {T.~P.}\ \bibnamefont {Cysne}}, \bibinfo {author} {\bibfnamefont {J.~H.}\ \bibnamefont {Garcia}}, \bibinfo {author} {\bibfnamefont {A.~R.}\ \bibnamefont {Rocha}},\ and\ \bibinfo {author} {\bibfnamefont {T.~G.}\ \bibnamefont {Rappoport}},\ }\bibfield  {title} {\bibinfo {title} {{Quantum Hall effect in graphene with interface-induced spin-orbit coupling}},\ }\href {https://doi.org/10.1103/PhysRevB.97.085413} {\bibfield  {journal} {\bibinfo  {journal} {Phys. Rev. B}\ }\textbf {\bibinfo {volume} {97}},\ \bibinfo {pages} {085413} (\bibinfo {year} {2018}{\natexlab{b}})}\BibitemShut {NoStop}%
\bibitem [{\citenamefont {Khatibi}\ and\ \citenamefont {Power}(2022)}]{Power-PhysRevB.106.125417}%
  \BibitemOpen
  \bibfield  {author} {\bibinfo {author} {\bibfnamefont {Z.}~\bibnamefont {Khatibi}}\ and\ \bibinfo {author} {\bibfnamefont {S.~R.}\ \bibnamefont {Power}},\ }\bibfield  {title} {\bibinfo {title} {{Proximity spin-orbit coupling in graphene on alloyed transition metal dichalcogenides}},\ }\href {https://doi.org/10.1103/PhysRevB.106.125417} {\bibfield  {journal} {\bibinfo  {journal} {Phys. Rev. B}\ }\textbf {\bibinfo {volume} {106}},\ \bibinfo {pages} {125417} (\bibinfo {year} {2022})}\BibitemShut {NoStop}%
\bibitem [{\citenamefont {Costa}\ \emph {et~al.}(2023)\citenamefont {Costa}, \citenamefont {Focassio}, \citenamefont {Canonico}, \citenamefont {Cysne}, \citenamefont {Schleder}, \citenamefont {Muniz}, \citenamefont {Fazzio},\ and\ \citenamefont {Rappoport}}]{Costa-PhysRevLett.130.116204}%
  \BibitemOpen
  \bibfield  {author} {\bibinfo {author} {\bibfnamefont {M.}~\bibnamefont {Costa}}, \bibinfo {author} {\bibfnamefont {B.}~\bibnamefont {Focassio}}, \bibinfo {author} {\bibfnamefont {L.~M.}\ \bibnamefont {Canonico}}, \bibinfo {author} {\bibfnamefont {T.~P.}\ \bibnamefont {Cysne}}, \bibinfo {author} {\bibfnamefont {G.~R.}\ \bibnamefont {Schleder}}, \bibinfo {author} {\bibfnamefont {R.~B.}\ \bibnamefont {Muniz}}, \bibinfo {author} {\bibfnamefont {A.}~\bibnamefont {Fazzio}},\ and\ \bibinfo {author} {\bibfnamefont {T.~G.}\ \bibnamefont {Rappoport}},\ }\bibfield  {title} {\bibinfo {title} {{Connecting Higher-Order Topology with the Orbital Hall Effect in Monolayers of Transition Metal Dichalcogenides}},\ }\href {https://doi.org/10.1103/PhysRevLett.130.116204} {\bibfield  {journal} {\bibinfo  {journal} {Phys. Rev. Lett.}\ }\textbf {\bibinfo {volume} {130}},\ \bibinfo {pages} {116204} (\bibinfo {year} {2023})}\BibitemShut {NoStop}%
\bibitem [{\citenamefont {Dai}\ \emph {et~al.}(2024)\citenamefont {Dai}, \citenamefont {Fu}, \citenamefont {Ang},\ and\ \citenamefont {Chen}}]{dai2024twodimensionalweylnodallinesemimetal}%
  \BibitemOpen
  \bibfield  {author} {\bibinfo {author} {\bibfnamefont {X.}~\bibnamefont {Dai}}, \bibinfo {author} {\bibfnamefont {P.-H.}\ \bibnamefont {Fu}}, \bibinfo {author} {\bibfnamefont {Y.~S.}\ \bibnamefont {Ang}},\ and\ \bibinfo {author} {\bibfnamefont {Q.}~\bibnamefont {Chen}},\ }\bibfield  {title} {\bibinfo {title} {Two-dimensional weyl nodal-line semimetal and antihelical edge states in a modified kane-mele model},\ }\href {https://doi.org/10.1103/PhysRevB.110.195409} {\bibfield  {journal} {\bibinfo  {journal} {Phys. Rev. B}\ }\textbf {\bibinfo {volume} {110}},\ \bibinfo {pages} {195409} (\bibinfo {year} {2024})}\BibitemShut {NoStop}%
\bibitem [{\citenamefont {Gmitra}\ \emph {et~al.}(2016)\citenamefont {Gmitra}, \citenamefont {Kochan}, \citenamefont {H\"ogl},\ and\ \citenamefont {Fabian}}]{Gmitra-PhysRevB.93.155104}%
  \BibitemOpen
  \bibfield  {author} {\bibinfo {author} {\bibfnamefont {M.}~\bibnamefont {Gmitra}}, \bibinfo {author} {\bibfnamefont {D.}~\bibnamefont {Kochan}}, \bibinfo {author} {\bibfnamefont {P.}~\bibnamefont {H\"ogl}},\ and\ \bibinfo {author} {\bibfnamefont {J.}~\bibnamefont {Fabian}},\ }\bibfield  {title} {\bibinfo {title} {{Trivial and inverted Dirac bands and the emergence of quantum spin Hall states in graphene on transition-metal dichalcogenides}},\ }\href {https://doi.org/10.1103/PhysRevB.93.155104} {\bibfield  {journal} {\bibinfo  {journal} {Phys. Rev. B}\ }\textbf {\bibinfo {volume} {93}},\ \bibinfo {pages} {155104} (\bibinfo {year} {2016})}\BibitemShut {NoStop}%
\bibitem [{\citenamefont {Bindel}\ \emph {et~al.}(2016)\citenamefont {Bindel}, \citenamefont {Pezzotta}, \citenamefont {Ulrich}, \citenamefont {Liebmann}, \citenamefont {Sherman},\ and\ \citenamefont {Morgenstern}}]{Bindel2016-NatPhys}%
  \BibitemOpen
  \bibfield  {author} {\bibinfo {author} {\bibfnamefont {J.~R.}\ \bibnamefont {Bindel}}, \bibinfo {author} {\bibfnamefont {M.}~\bibnamefont {Pezzotta}}, \bibinfo {author} {\bibfnamefont {J.}~\bibnamefont {Ulrich}}, \bibinfo {author} {\bibfnamefont {M.}~\bibnamefont {Liebmann}}, \bibinfo {author} {\bibfnamefont {E.~Y.}\ \bibnamefont {Sherman}},\ and\ \bibinfo {author} {\bibfnamefont {M.}~\bibnamefont {Morgenstern}},\ }\bibfield  {title} {\bibinfo {title} {{Probing variations of the Rashba spin–orbit coupling at the nanometre scale}},\ }\href {https://doi.org/10.1038/nphys3774} {\bibfield  {journal} {\bibinfo  {journal} {Nature Physics}\ }\textbf {\bibinfo {volume} {12}},\ \bibinfo {pages} {920–925} (\bibinfo {year} {2016})}\BibitemShut {NoStop}%
\bibitem [{\citenamefont {Str\"om}\ \emph {et~al.}(2010)\citenamefont {Str\"om}, \citenamefont {Johannesson},\ and\ \citenamefont {Japaridze}}]{Strom-PhysRevLett.104.256804}%
  \BibitemOpen
  \bibfield  {author} {\bibinfo {author} {\bibfnamefont {A.}~\bibnamefont {Str\"om}}, \bibinfo {author} {\bibfnamefont {H.}~\bibnamefont {Johannesson}},\ and\ \bibinfo {author} {\bibfnamefont {G.~I.}\ \bibnamefont {Japaridze}},\ }\bibfield  {title} {\bibinfo {title} {{Edge Dynamics in a Quantum Spin Hall State: Effects from Rashba Spin-Orbit Interaction}},\ }\href {https://doi.org/10.1103/PhysRevLett.104.256804} {\bibfield  {journal} {\bibinfo  {journal} {Phys. Rev. Lett.}\ }\textbf {\bibinfo {volume} {104}},\ \bibinfo {pages} {256804} (\bibinfo {year} {2010})}\BibitemShut {NoStop}%
\bibitem [{\citenamefont {Aligia}(2023)}]{Aligia-PhysRevB.107.075153}%
  \BibitemOpen
  \bibfield  {author} {\bibinfo {author} {\bibfnamefont {A.~A.}\ \bibnamefont {Aligia}},\ }\bibfield  {title} {\bibinfo {title} {Topological invariants based on generalized position operators and application to the interacting rice-mele model},\ }\href {https://doi.org/10.1103/PhysRevB.107.075153} {\bibfield  {journal} {\bibinfo  {journal} {Phys. Rev. B}\ }\textbf {\bibinfo {volume} {107}},\ \bibinfo {pages} {075153} (\bibinfo {year} {2023})}\BibitemShut {NoStop}%
\bibitem [{\citenamefont {Aligia}\ and\ \citenamefont {Ortiz}(1999)}]{Aligia-PhysRevLett.82.2560}%
  \BibitemOpen
  \bibfield  {author} {\bibinfo {author} {\bibfnamefont {A.~A.}\ \bibnamefont {Aligia}}\ and\ \bibinfo {author} {\bibfnamefont {G.}~\bibnamefont {Ortiz}},\ }\bibfield  {title} {\bibinfo {title} {Quantum mechanical position operator and localization in extended systems},\ }\href {https://doi.org/10.1103/PhysRevLett.82.2560} {\bibfield  {journal} {\bibinfo  {journal} {Phys. Rev. Lett.}\ }\textbf {\bibinfo {volume} {82}},\ \bibinfo {pages} {2560} (\bibinfo {year} {1999})}\BibitemShut {NoStop}%
\bibitem [{\citenamefont {Duan}\ \emph {et~al.}(2014)\citenamefont {Duan}, \citenamefont {Wang}, \citenamefont {Shaw}, \citenamefont {Cheng}, \citenamefont {Chen}, \citenamefont {Li}, \citenamefont {Wu}, \citenamefont {Tang}, \citenamefont {Zhang}, \citenamefont {Pan}, \citenamefont {Jiang}, \citenamefont {Yu}, \citenamefont {Huang},\ and\ \citenamefont {Duan}}]{Lateral-HeterostructureDuan2014}%
  \BibitemOpen
  \bibfield  {author} {\bibinfo {author} {\bibfnamefont {X.}~\bibnamefont {Duan}}, \bibinfo {author} {\bibfnamefont {C.}~\bibnamefont {Wang}}, \bibinfo {author} {\bibfnamefont {J.~C.}\ \bibnamefont {Shaw}}, \bibinfo {author} {\bibfnamefont {R.}~\bibnamefont {Cheng}}, \bibinfo {author} {\bibfnamefont {Y.}~\bibnamefont {Chen}}, \bibinfo {author} {\bibfnamefont {H.}~\bibnamefont {Li}}, \bibinfo {author} {\bibfnamefont {X.}~\bibnamefont {Wu}}, \bibinfo {author} {\bibfnamefont {Y.}~\bibnamefont {Tang}}, \bibinfo {author} {\bibfnamefont {Q.}~\bibnamefont {Zhang}}, \bibinfo {author} {\bibfnamefont {A.}~\bibnamefont {Pan}}, \bibinfo {author} {\bibfnamefont {J.}~\bibnamefont {Jiang}}, \bibinfo {author} {\bibfnamefont {R.}~\bibnamefont {Yu}}, \bibinfo {author} {\bibfnamefont {Y.}~\bibnamefont {Huang}},\ and\ \bibinfo {author} {\bibfnamefont {X.}~\bibnamefont {Duan}},\ }\bibfield  {title} {\bibinfo {title} {Lateral epitaxial growth of two-dimensional layered semiconductor heterojunctions},\ }\href
  {https://doi.org/10.1038/nnano.2014.222} {\bibfield  {journal} {\bibinfo  {journal} {Nature Nanotechnology}\ }\textbf {\bibinfo {volume} {9}},\ \bibinfo {pages} {1024–1030} (\bibinfo {year} {2014})}\BibitemShut {NoStop}%
\bibitem [{\citenamefont {Bianco}\ and\ \citenamefont {Resta}(2013)}]{Resta-Bianco-PhysRevLett.110.087202}%
  \BibitemOpen
  \bibfield  {author} {\bibinfo {author} {\bibfnamefont {R.}~\bibnamefont {Bianco}}\ and\ \bibinfo {author} {\bibfnamefont {R.}~\bibnamefont {Resta}},\ }\bibfield  {title} {\bibinfo {title} {Orbital magnetization as a local property},\ }\href {https://doi.org/10.1103/PhysRevLett.110.087202} {\bibfield  {journal} {\bibinfo  {journal} {Phys. Rev. Lett.}\ }\textbf {\bibinfo {volume} {110}},\ \bibinfo {pages} {087202} (\bibinfo {year} {2013})}\BibitemShut {NoStop}%
\bibitem [{\citenamefont {Marrazzo}\ and\ \citenamefont {Resta}(2016)}]{Marrazzo2016-PRL116}%
  \BibitemOpen
  \bibfield  {author} {\bibinfo {author} {\bibfnamefont {A.}~\bibnamefont {Marrazzo}}\ and\ \bibinfo {author} {\bibfnamefont {R.}~\bibnamefont {Resta}},\ }\bibfield  {title} {\bibinfo {title} {Irrelevance of the boundary on the magnetization of metals},\ }\href {https://doi.org/10.1103/PhysRevLett.116.137201} {\bibfield  {journal} {\bibinfo  {journal} {Phys. Rev. Lett.}\ }\textbf {\bibinfo {volume} {116}},\ \bibinfo {pages} {137201} (\bibinfo {year} {2016})}\BibitemShut {NoStop}%
\bibitem [{\citenamefont {Bianco}\ and\ \citenamefont {Resta}(2016)}]{Bianco2016-PRB93}%
  \BibitemOpen
  \bibfield  {author} {\bibinfo {author} {\bibfnamefont {R.}~\bibnamefont {Bianco}}\ and\ \bibinfo {author} {\bibfnamefont {R.}~\bibnamefont {Resta}},\ }\bibfield  {title} {\bibinfo {title} {Orbital magnetization in insulators: Bulk versus surface},\ }\href {https://doi.org/10.1103/PhysRevB.93.174417} {\bibfield  {journal} {\bibinfo  {journal} {Phys. Rev. B}\ }\textbf {\bibinfo {volume} {93}},\ \bibinfo {pages} {174417} (\bibinfo {year} {2016})}\BibitemShut {NoStop}%
\bibitem [{\citenamefont {Drigo}\ and\ \citenamefont {Resta}(2020)}]{Drigo2020-PRB101}%
  \BibitemOpen
  \bibfield  {author} {\bibinfo {author} {\bibfnamefont {E.}~\bibnamefont {Drigo}}\ and\ \bibinfo {author} {\bibfnamefont {R.}~\bibnamefont {Resta}},\ }\bibfield  {title} {\bibinfo {title} {Chern number and orbital magnetization in ribbons, polymers, and single-layer materials},\ }\href {https://doi.org/10.1103/PhysRevB.101.165120} {\bibfield  {journal} {\bibinfo  {journal} {Phys. Rev. B}\ }\textbf {\bibinfo {volume} {101}},\ \bibinfo {pages} {165120} (\bibinfo {year} {2020})}\BibitemShut {NoStop}%
\bibitem [{\citenamefont {Seleznev}\ and\ \citenamefont {Vanderbilt}(2023)}]{Seleznev-Vanderbilt-PhysRevB.107.115102}%
  \BibitemOpen
  \bibfield  {author} {\bibinfo {author} {\bibfnamefont {D.}~\bibnamefont {Seleznev}}\ and\ \bibinfo {author} {\bibfnamefont {D.}~\bibnamefont {Vanderbilt}},\ }\bibfield  {title} {\bibinfo {title} {Towards a theory of surface orbital magnetization},\ }\href {https://doi.org/10.1103/PhysRevB.107.115102} {\bibfield  {journal} {\bibinfo  {journal} {Phys. Rev. B}\ }\textbf {\bibinfo {volume} {107}},\ \bibinfo {pages} {115102} (\bibinfo {year} {2023})}\BibitemShut {NoStop}%
\bibitem [{\citenamefont {Marrazzo}\ and\ \citenamefont {Resta}(2017)}]{Marrazzo2017-PRB}%
  \BibitemOpen
  \bibfield  {author} {\bibinfo {author} {\bibfnamefont {A.}~\bibnamefont {Marrazzo}}\ and\ \bibinfo {author} {\bibfnamefont {R.}~\bibnamefont {Resta}},\ }\bibfield  {title} {\bibinfo {title} {Locality of the anomalous hall conductivity},\ }\href {https://doi.org/10.1103/PhysRevB.95.121114} {\bibfield  {journal} {\bibinfo  {journal} {Phys. Rev. B}\ }\textbf {\bibinfo {volume} {95}},\ \bibinfo {pages} {121114} (\bibinfo {year} {2017})}\BibitemShut {NoStop}%
\bibitem [{\citenamefont {Rauch}\ \emph {et~al.}(2018)\citenamefont {Rauch}, \citenamefont {Olsen}, \citenamefont {Vanderbilt},\ and\ \citenamefont {Souza}}]{AHC-PhysRevB.98.115108}%
  \BibitemOpen
  \bibfield  {author} {\bibinfo {author} {\bibfnamefont {T.~c.~v.}\ \bibnamefont {Rauch}}, \bibinfo {author} {\bibfnamefont {T.}~\bibnamefont {Olsen}}, \bibinfo {author} {\bibfnamefont {D.}~\bibnamefont {Vanderbilt}},\ and\ \bibinfo {author} {\bibfnamefont {I.}~\bibnamefont {Souza}},\ }\bibfield  {title} {\bibinfo {title} {Geometric and nongeometric contributions to the surface anomalous hall conductivity},\ }\href {https://doi.org/10.1103/PhysRevB.98.115108} {\bibfield  {journal} {\bibinfo  {journal} {Phys. Rev. B}\ }\textbf {\bibinfo {volume} {98}},\ \bibinfo {pages} {115108} (\bibinfo {year} {2018})}\BibitemShut {NoStop}%
\bibitem [{\citenamefont {Rauch}\ \emph {et~al.}(2020)\citenamefont {Rauch}, \citenamefont {T\"opler},\ and\ \citenamefont {Mertig}}]{Mertig-PhysRevB.101.064206}%
  \BibitemOpen
  \bibfield  {author} {\bibinfo {author} {\bibfnamefont {T.~c.~v.}\ \bibnamefont {Rauch}}, \bibinfo {author} {\bibfnamefont {F.}~\bibnamefont {T\"opler}},\ and\ \bibinfo {author} {\bibfnamefont {I.}~\bibnamefont {Mertig}},\ }\bibfield  {title} {\bibinfo {title} {Local spin hall conductivity},\ }\href {https://doi.org/10.1103/PhysRevB.101.064206} {\bibfield  {journal} {\bibinfo  {journal} {Phys. Rev. B}\ }\textbf {\bibinfo {volume} {101}},\ \bibinfo {pages} {064206} (\bibinfo {year} {2020})}\BibitemShut {NoStop}%
\end{thebibliography}%

\end{document}